\documentclass[12pt]{article}
\usepackage{graphics,graphicx,graphpap,color}
\usepackage{setspace}

\newcommand{\GeV}{~\mbox{GeV}}
\newcommand{\TeV}{~\mbox{TeV}}

\newcommand\bmat{\left( \begin{array}{cc}}
\newcommand\emat{\end{array}\right)}

\def\msbar{\ifmmode{\overline{\rm MS}} \else{$\overline{\rm MS}$} \fi}
\def\drbar{\ifmmode{\overline{\rm DR}} \else{$\overline{\rm DR}$} \fi}

\def\ti              {\tilde}

\def\a               {\alpha}
\def\b               {\beta}
\def\d               {\delta}
\def\D               {\Delta}

\def\g               {\gamma}
\def\G               {\Gamma}
\def\l               {\lambda}
\def\t               {\theta}

\def\x               {\chi}

\renewcommand{\sf}{{\tilde{f}}}

\newcommand{\st}{{\tilde{t}}}
\renewcommand{\sb}{{\tilde{b}}}
\newcommand{\stau}{{\tilde{\tau}}}
\newcommand{\snutau}{{{\tilde{\nu}_\tau}}}
\newcommand{\su}{{\tilde{u}}}
\newcommand{\sd}{{\tilde{d}}}
\newcommand{\sen}{{{\tilde{\nu}_e}}}
\newcommand{\se}{{\tilde e}}
\renewcommand{\sc}{{\tilde{c}}}
\newcommand{\sstrange}{{\tilde{s}}}
\newcommand{\smn}{{{\tilde{\nu}_\mu}}}
\newcommand{\sm}{{\tilde \mu}}

\def\sF{\ti F}

\newcommand{\Rst}{R^{\ti t}}
\newcommand{\Rsb}{R^{\ti b}}
\newcommand{\Rsl}{R^{\tilde{\tau}}}

\newcommand{\Rsu}{R^{\ti u}}
\newcommand{\Rsd}{R^{\ti d}}
\newcommand{\Rse}{R^{\tilde{e}}}

\newcommand{\Rsc}{R^{\ti c}}
\newcommand{\Rss}{R^{\ti s}}
\newcommand{\Rsm}{R^{\tilde{\mu}}}

\newcommand{\cha}[1]   {{\ti \x^+_{#1}}}

\newcommand{\neu}[1]   {{\ti \x^0_{#1}}}

\def\chp             {\ti \x^+}

\def\nt              {\ti \x^0}

\newcommand{\msf}[1]   {m_{\ti f_{#1}}}

\def\tw              {\t_{W}}

\def\onehfb          {\frac{1}{2}}

\def\vor             {\frac{1}{(4\pi)^2}}

\def\non             {\nonumber}

\renewcommand\Re{{\rm Re}}

\renewcommand\d{\delta}

\begin{document}

\pagestyle{empty} \vspace*{-1cm}

\begin{flushright}
  HEPHY-PUB 832/07 \\
  hep-ph/0701134
\end{flushright}

\vspace*{2cm}

\begin{center}
{\Large\bf\boldmath
   Complete one-loop corrections to decays of
   \\[2mm]
   charged and CP-even neutral Higgs bosons 
   \\[3mm] 
   into sfermions} \\[5mm]

\vspace{10mm}

$\mbox{C.~Weber}, \mbox{K.~Kova\v{r}\'{\i}k},
\mbox{H.~Eberl}, \mbox{W.~Majerotto}$\\[5mm]

\vspace{6mm} $$\mbox{{\it Institut f\"ur Hochenergiephysik der 
\"Osterreichischen Akademie der Wissenschaften,}}$$\vspace{-0.9cm} 
$$\mbox{{\it A-1050 Vienna, Austria}}$$ 
\end{center}

\vspace{20mm}

\begin{abstract} \noindent We present the full one-loop corrections to 
charged and CP-even neutral Higgs boson decays into sfermions including 
also the crossed channels. The calculation was carried out in the 
minimal supersymmetric extension of the Standard Model and we use the 
on-shell renormalization scheme. For the down-type sfermions, we use 
\drbar running fermion masses and the trilinear coupling $A_f$ as 
input. Furthermore, we present the first numerical analysis for decays 
according to the Supersymmetric Parameter Analysis project. This 
requires the renormalization of the whole MSSM. The corrections are 
found to be numerically stable and not negligible. \end{abstract} 

\vfill
\newpage
\pagestyle{plain} \setcounter{page}{2}

\section{Introduction}
The Higgs boson is the last not discovered particle of the Standard 
Model (SM) and so the search for the Higgs boson is the prime objective 
of the LHC and other future colliders. Apart from the SM, the Higgs 
boson is also predicted by its minimal supersymmetric extension - the 
Minimal Supersymmetric Standard Model (MSSM). As opposed to the SM, the 
MSSM has not only one neutral Higgs boson but it predicts the existence 
of two neutral CP-even Higgs bosons ($h^0$, $H^0$), one neutral CP-odd 
Higgs boson ($A^0$) and two charged Higgs bosons ($H^\pm$). The 
existence of a charged Higgs boson or a CP-odd neutral one would be 
clear evidence for physics beyond the SM. 
\newline %
A further difference in the MSSM is the possibility for the Higgs 
bosons to decay not only into SM particles. In case the supersymmetric 
(SUSY) partners are not too heavy, the Higgs bosons can decay into SUSY 
particles as well (neutralinos $\neu{i}$, charginos $\cha{k}$ and 
sfermions $\sf_m$). The new decay channels might substantially 
influence the branching ratios of the MSSM Higgs bosons. 
\newline %
At tree-level the decays into SUSY particles were studied in 
\cite{tree1,tree2} and one-loop effects of the decays into charginos 
and neutralinos were analyzed in \cite{0111303,Zhang} and were found 
not to be negligible. For the case of the CP-odd Higgs boson also the 
full one-loop corrections to the decay into sfermions were analyzed in 
\cite{A0sq_letter,A0paper}. 
\newline %
This paper is the continuation of the effort in 
\cite{A0sq_letter,A0paper} and includes the decays of the remaining 
Higgs bosons of the MSSM into sfermions  (including crossed channels 
$\sf_2 \rightarrow \sf_1 h^0$). It also extends the SUSY-QCD one-loop 
analysis of \cite{SUSY-QCD} by including all SUSY-QCD and electroweak 
effects. The emphasis is put on the decay into 3rd generation sfermions 
as their masses can be light due to large mixings. Nevertheless, 
analytical and numerical results are presented for all generations of 
sfermions (i.e. $h_k^0\rightarrow \sf_i\ {\bar{\!\!\tilde{f}}}_{\!j}$ 
and $H^\pm\rightarrow \sf_i\bar{\sf'}_{\!\!\!j}$ where $h^0_k = 
(h^0,H^0)$ and $\sf = 
(\tilde{u},\tilde{d},\tilde{s},\tilde{c},\tilde{b}, 
\tilde{t},\tilde{e},\tilde{\mu},\tilde{\tau}$). 
\newline %
The full electroweak corrections are calculated in the on-shell scheme 
\cite{onshellren} in the MSSM with real parameters. Due to the known 
problems of the on-shell scheme as demonstrated in \cite{A0paper}, the 
artificially large on-shell parameters are replaced by the 
corresponding \drbar counterparts. The numerical analysis is made using 
the \drbar input defined by the Supersymmetric Parameter Analysis 
Project (SPA) \cite{SPA}. In contrast to \cite{A0paper}, the actual 
calculation uses an on-shell input set fully consistent with the SPA 
convention. In order to obtain such a input set, the renormalization of 
the whole MSSM is required. 
\newline %
The paper is organized as follows. In section 2 the tree-level formulae 
are given for all decays. Section 3 and 4 show the full electroweak 
corrections including the bremsstrahlung using the analytic formulae 
from the appendices A and B. The numerical analysis is presented in 
section 5 and section 6 summarizes our conclusions. 

\section{Tree-level result}\label{treelevel}
The tree-level widths for a neutral Higgs $h_{\{1,2\}}^0 = \{h^0,
H^0\}$ decaying into two scalar fermions, $h_k^0 \rightarrow
\tilde{f}_i \ {\bar{\!\!\tilde{f}}}_{\!j}$ with $i, j = (1, 2)$, are 
given by 
\begin{eqnarray}
\G^{\rm tree}(h_k^0 \rightarrow \tilde{f}_i \
{\bar{\!\!\tilde{f}}}_{\!j}) &=& \frac{N_C^f\, \kappa
(m_{h_k^0}^2, m^2_{\sf_i}, m^2_{\sf_j})}{16 \,\pi\, m^3_{h_k^0}}\
|G_{ijk}^{\sf}|^2
\end{eqnarray}
with $\kappa (x, y, z) = \sqrt{(x-y-z)^2 - 4 y z}$ and the colour
factor $N_C^f = 3$ for squarks and \mbox{$N_C^f = 1$} for
sleptons, respectively. \\
Analogously, the decay width for the charged Higgs boson $H^+$ is given by
\begin{eqnarray}
\G^{\rm tree} (H^+\rightarrow \sf^{\uparrow}_i \,\bar{\sf^{\downarrow}_{j}}) & = & 
\frac{N_C^f\,\kappa(m^2_{H^+}, m_{\sf^{\uparrow}_i}^2, m^2_{\sf^{\downarrow}_j})}{16\pi 
m_{H^+}^3}|G_{ij1}^{\uparrow\downarrow}|^2\,, 
\end{eqnarray}
where $\sf^{\uparrow/\downarrow}$ stand for the up-type or down-type 
sfermions. The sfermion-Higgs boson couplings $G_{ijk}^{\sf}$ and 
$G_{ij1}^{\uparrow\downarrow}$, defined by the interaction lagrangian 
${\cal L}_{\rm int} = G_{ijk}^{\sf}\, h_k^0 \sf_i^\ast \sf_j + 
G_{ij1}^{\uparrow\downarrow}\, H^+ \sf^{\uparrow\ast}_i 
\sf^{\downarrow}_j$ as well as all couplings needed in this paper, are 
given in \cite{A0paper}. 

The sfermion mass matrix is diagonalized by a
real 2\,x\,2 rotation matrix $R^\sf_{i\a}$ with rotation angle
$\theta_{\!\sf}$ \cite{GunionHaber1, GunionHaber2},
\begin{eqnarray}
  {\cal M}_{\sf}^{\,2} \,=\,
   \left(
     \begin{array}{cc}
       m_{LL}^{\,2} & m_{LR}^{\,2}
       \\[2mm]
       m_{RL}^{\,2} & m_{RR}^{\,2}
     \end{array}
   \right)
  =
   \left(
     \begin{array}{cc}
       m_{\sf_L}^{\,2} & a_f\, m_f
       \\[2mm]
       a_f\,m_f & m_{\sf_R}^{\,2}
     \end{array}
   \right)
  = \left( R^\sf \right)^\dag
   \left(
     \begin{array}{cc}
       m_{\sf_1}^{\,2} & 0
       \\[2mm]
       0 & m_{\sf_2}^{\,2}
     \end{array}
   \right) R^\sf \,,
\end{eqnarray}
which relates the mass eigenstates $\sf_i$, $i = 1, 2$, $(m_{\sf_1} < 
m_{\sf_2})$ to the gauge eigenstates $\sf_\a$, $\a = L, R$, by $\sf_i = 
R^\sf_{i\a} \sf_\a$. The left- and right-handed and the left-right 
mixing entries in the mass matrix are given by 
\begin{eqnarray}
  m_{\sf_L}^{\,2} &=& M_{\{\ti Q\!,\,\ti L \}}^2
       + (I^{3L}_f \!-\! e_{f}^{}\sin^2\!\tw)\cos2\b\,
       m_{Z}^{\,2}
       + m_{f}^2\,, \\[2mm]\label{MsD}
  m_{\sf_R}^{\,2} &=& M_{\{\ti U\!,\,\ti D\!,\,\ti E \}}^2
       + e_{f}\sin^2\!\tw \cos2\b\,m_{Z}^{\,2}
       + m_f^2\,, \\[2mm]
  a_f &=& A_f - \mu \,(\tan\b)^{-2 I^{3L}_f} \,.
\end{eqnarray} $M_{\ti Q}$, $M_{\ti L}$, $M_{\ti U}$, $M_{\ti D}$ and 
$M_{\ti E}$ are soft SUSY breaking masses, $A_f$ is the trilinear 
scalar coupling parameter, $\mu$ the higgsino mass parameter, $\tan\b = 
\frac{v_2}{v_1}$ is the ratio of the vacuum expectation values of the 
two neutral Higgs doublet states \cite{GunionHaber1,GunionHaber2}, 
$I^{3L}_f$ denotes the third component of the weak isospin of the 
left-handed fermion $f$, $e_f$ the electric charge in terms of the 
elementary charge $e_0$, and $\tw$ is the Weinberg angle. \\ The mass 
eigenvalues and the mixing angle in terms of primary parameters are 
\begin{eqnarray} 
  \msf{1,2}^2
    &=& \frac{1}{2} \left(
    \msf{L}^2 + \msf{R}^2 \mp
    \sqrt{(\msf{L}^2 \!-\! \msf{R}^2)^2 + 4 a_f^2 m_f^2}\,\right) \,,
\\
  \cos\t_{\sf}
    &=& \frac{-a_f\,m_f}
    {\sqrt{(\msf{L}^2 \!-\! \msf{1}^2)^2 + a_f^2 m_f^2}}
  \hspace{2cm} (0\leq \t_{\sf} < \pi) \,,
\end{eqnarray} and the trilinear soft breaking parameter $A_f$ can be 
written as 
\begin{eqnarray}\label{mFaF} 
A_f &=& \frac{1}{2m_f} \left(m_{\sf_1}^2-m_{\sf_2}^2 \right) \sin 2\theta_\sf 
+ \mu \,(\tan\b)^{-2 I^{3L}_f} \,. 
\end{eqnarray} The mass of the sneutrino 
$\snutau$ is given by $m_{\snutau}^2 = M_{\ti L}^2 + \frac{1}{2}\,m_Z^2 
\cos2\beta$. \\ For the crossed channels, $\sf_2 \rightarrow \sf_1 
h_k^0$ and $\sf^\uparrow_2 \rightarrow \sf^\downarrow_1 H^+$, the decay 
widths are \begin{eqnarray} \G^{\rm tree}(\tilde{f}_2 \rightarrow 
\tilde{f}_1 h_k^0) &=& \frac{\kappa (m_{h_k^0}^2, m^2_{\sf_1}, 
m^2_{\sf_2})}{16 \,\pi\, m^3_{\sf_2}}\ |G_{12k}^{\sf}|^2 \,, \\ \G^{\rm 
tree}(\sf^\uparrow_j \rightarrow \sf^\downarrow_i H^+) &=& 
\frac{\kappa(m^2_{H^+}, m_{\sf^{\uparrow}_j}^2, 
m^2_{\sf^{\downarrow}_i})}{16 \,\pi\, m^3_{\sf^\uparrow_j}}\ 
|G_{ij1}^{\uparrow\downarrow}|^2\,. \end{eqnarray} \vspace{2mm} 
\section{One-loop Corrections}\label{renormalization}
The one-loop corrected (renormalized) amplitudes $G_{ijk}^{\sf\, \rm 
ren}$ and $G_{ij1}^{\uparrow\downarrow, \rm ren}$ can be expressed as 
\begin{eqnarray}
G_{ijk}^{\sf, \rm ren} &=& G_{ijk}^{\sf} + \D G_{ijk}^{\sf} ~=~ 
G_{ijk}^{\sf} + \d G_{ijk}^{\sf (v)} + \d G_{ijk}^{\sf (w)} + \d 
G_{ijk}^{\sf (c)} \,,
\\[2mm]
G_{ij1}^{\uparrow\downarrow, \rm ren} &=& G_{ij1}^{\uparrow\downarrow} + \D G_{ij1}^{\uparrow\downarrow} 
~=~ G_{ij1}^{\uparrow\downarrow} + \d G_{ij1}^{\uparrow\downarrow (v)} + \d G_{ij1}^{\uparrow\downarrow 
(w)} + \d G_{ij1}^{\uparrow\downarrow (c)} \,,
\end{eqnarray}
where $\d G_{ijk}^{\sf (v)}, \d G_{ijk}^{\sf (w)}$ and $\d G_{ijk}^{\sf 
(c)}$ and the corresponding terms for the couplings to the charged 
Higgs boson stand for the vertex corrections, the wave-function 
corrections and the coupling counter-term corrections due to the shifts 
from the bare to the on-shell values, respectively. 
\\%
Throughout the paper we use the SUSY invariant dimensional reduction 
$(\overline{\rm DR})$ as a regularization scheme. For convenience we 
perform the calculation in the 't Hooft-Feynman gauge, $\xi=1$. 
\\%
The vertex corrections $\d G_{ijk}^{\sf (v)}$ and $\d 
G_{ij1}^{\uparrow\downarrow (v)}$ come from the diagrams listed in 
Figs.~\ref{vertex-graphshk0} and \ref{Hpvertex-graphs}. The analytic 
formulae are given in Appendix \ref{appVertex}. The wave-function 
corrections $\d G_{ijk}^{\sf (w)}$ can be written as 
\begin{eqnarray}
\delta G_{ijk}^{\sf (w)} = \frac{1}{2}\left[\delta Z_{i'\!i}^\sf
\, G_{i'\!jk}^\sf + \delta Z_{j'\!j}^\sf \, G_{ij'\!k}^\sf +
\delta Z_{lk}^H \, G_{ijl}^\sf\right] \,,
\end{eqnarray}
with the implicit summations over $i', j', l = 1, 2$. The wave-function 
renormalization constants are determined by imposing the on-shell 
renormalization conditions \cite{onshellren} 
\begin{eqnarray}
   &&\begin{array}{l@{\qquad\qquad}l}
      \delta Z_{ii}^{\sf} ~=~ - \Re\,\dot\Pi_{ii}^{\sf} (m_{\sf_i}^2)\,,
      & i = 1,2\,,
      \\[3mm]
      \delta Z_{ij}^{\sf} ~=~ \displaystyle{\frac{2}{m_{\sf_i}^2\!-
      \!m_{\sf_j}^2}}\, \Re\, \Pi_{ij}^{\sf}(m_{\sf_j}^2) \,,
      & i,j = (1,2),\ i \neq j,\ \sf \neq \ti\nu_{e,\mu,\tau}
   \end{array}
\\[2mm]
   &&\begin{array}{l@{\qquad\qquad}l}
      \delta Z_{kk}^{H} ~=~ - \Re\,\dot\Pi_{kk}^{H} (m_{h_k^0}^2)\,,
      & k = 1,2\,,
      \\[3mm]
      \delta Z_{kl}^{H} ~=~ \displaystyle{\frac{2}{m_{h_k^0}^2\!-
      \!m_{h_l^0}^2}}\, \Re\, \Pi_{kl}^{H}(m_{h_l^0}^2) \,,
      & k,l = (1,2),\ k \neq l\,.
   \end{array}
\end{eqnarray}
The explicit forms of the off-diagonal Higgs boson and sfermion 
self-energies and their derivatives, $\Pi_{kl}^{H}$, $\dot\Pi_{kk}^{H}$ 
and $\Pi_{ij}^{\sf}$, $\dot\Pi_{ii}^\sf$ are given in Appendix 
\ref{selfenergies} and in \cite{A0paper}. 
\newline %
The coupling counter-term corrections which come from the shifting of 
the parameters in the lagrangian can be expressed as 
\begin{eqnarray}\label{dGsfijkc}
\delta G_{ijk}^{\sf (c)} & = & \left[\delta R^\sf\cdot
G_{LR,k}^\sf\cdot (R^\sf)^T + R^\sf \cdot \d G_{LR,k}^\sf \cdot(
R^\sf)^T + \ R^\sf\cdot G_{LR,k}^\sf\cdot (\d R^\sf)^T
\right]_{ij} \,.
\end{eqnarray}
The counter term for the sfermion mixing angle, $\d \theta_{\!\sf}$, is 
determined such as to cancel the anti-hermitian part of the sfermion 
wave-function corrections \cite{guasch, JHEP9905}. Analogously we fix 
the Higgs boson mixing angle $\a$ by means of 
\begin{eqnarray}\label{dthetaalpha}
   \delta \a & = & \frac{1}{4}\, \Big(
   \d Z^H_{21} - \d Z^H_{12}\Big)
   = \frac{1}{2\big(m_{H^0}^2 \!-\! m_{h^0}^2\big)}\, {\Re}
   \left( \Pi_{12}^H(m_{H^0}^2) + \Pi_{21}^H(m_{h^0}^2)
    \right) \,.
\end{eqnarray}
Using the relations
\begin{eqnarray}
\frac{\delta G_{ij1}^\sf}{\delta \alpha} \ = \ -  G_{ij2}^\sf\,,
\qquad\quad \frac{\delta G_{ij2}^\sf}{\delta \alpha} \ = \
G_{ij1}^\sf \,,
\end{eqnarray}
and absorbing the counter terms for the mixing angles of the outer
particles, $\d\a$ and $\d\theta_{\!\sf}$, into $\delta G_{ijk}^{\sf 
(w)}$ yields the symmetric wave-function corrections 
\begin{eqnarray}\non\label{symmWF}
\delta G_{ijk}^{\sf (w,\,\rm symm.)} & = & \frac{1}{4} \Big(\delta
Z^\sf_{ii'} + \delta Z^\sf_{i'\!i}\Big) G_{i'\!jk}^\sf +
\frac{1}{4} \Big(\delta Z^\sf_{jj'} + \delta Z^\sf_{j'\!j}\Big)
G_{ij'\!k}^\sf +\frac{1}{4} \Big(\delta Z_{kl}^H + \delta
Z_{lk}^H\Big) G_{ijl}^\sf \,.
\\
\end{eqnarray}
Note that in this symmetrized form momentum-independent contributions 
from four-scalar couplings and tadpole shifts cancel out. 
\\ %
The sum of wave-function and counter-term corrections then reads
\begin{eqnarray}
\d G_{ijk}^{\sf (w+c)} & = & \d G_{ijk}^{\sf (w,\,\rm symm.)} + \Big[
R^\sf \cdot \hat\d G_{LR,k}^{\sf} \cdot ( R^\sf)^T \Big]_{ij} \,, 
\end{eqnarray}
The explicit forms of the counter terms $\hat\d G_{LR,k}^{\sf}$ for
$k=1, 2$ are given by 
\begin{eqnarray}\non
\big( \hat\d G_{LR,1}^\sf \big)_{11} & = & -\sqrt 2\, h_f\, m_f \,c_\a
\bigg( \frac{\d h_f}{h_f} + \frac{\d m_f}{m_f} \bigg) - g_{Z}\, m_{Z}\,
e_f\, \d s_{W}^2\, s_{\a+\b} 
\\[2mm]
&& \label{eqlist1} + g_Z\, m_Z (I_f^{3L}\!-\!e_f s_{W}^2) s_{\a+\b} \bigg( \frac{\d
g_Z}{g_Z} + \frac{\d m_Z}{m_Z} + \frac{\d\b}{t_{\a+\b}} \bigg) \,,
\\[2mm]
\big( \hat\d G_{LR,1}^\sf \big)_{12} & = & \frac{\d h_f}{h_f}
\big(G_{LR,1}^\sf \big)_{12} -  \frac{h_f}{\sqrt 2}\, (\d A_f \,c_\a +
\d\mu\, s_\a) \,, 
\\[2mm]\non
\big( \hat\d G_{LR,1}^\sf \big)_{22} & = & -\sqrt 2\, h_f\, m_f \,c_\a
\bigg( \frac{\d  h_f}{h_f} + \frac{\d m_f}{m_f} \bigg)
\\[2mm]
&& + g_Z\, m_Z \,e_f \,s_{W}^2\, s_{\a+\b} \bigg( \frac{\d g_Z}{g_Z} +
\frac{\d  m_Z}{m_Z} + \frac{\d s_W^2}{s_W^2} + \frac{\d\b}{t_{\a+\b}}
\bigg) 
\end{eqnarray}
for the sfermion couplings to the Higgs boson $h^0$ and
\begin{eqnarray}\non
\big( \hat\d G_{LR,2}^\sf \big)_{11} & = & -\sqrt 2\, h_f\, m_f \,s_\a 
\bigg( \frac{\d h_f}{h_f} + \frac{\d m_f}{m_f} \bigg) + g_{Z}\, m_{Z}\,
e_f\, \d s_{W}^2\, c_{\a+\b}
\\[2mm]
&& - g_Z\, m_Z (I_f^{3L}\!-\!e_f s_{W}^2) c_{\a+\b} \bigg( \frac{\d g_Z}{g_Z} +
\frac{\d m_Z}{m_Z} - t_{\a+\b}\, \d\b \bigg) \,,
\\[2mm]
\big( \hat\d G_{LR,2}^\sf \big)_{12} & = & \frac{\d h_f}{h_f}
\big(G_{LR,2}^\sf \big)_{12} - \frac{h_f}{\sqrt 2}\, (\d A_f \,s_\a -
\d\mu\, c_\a) \,,
\\[2mm]\non
\big( \hat\d G_{LR,2}^\sf \big)_{22} & = & -\sqrt 2\, h_f\, m_f \,s_\a 
\bigg( \frac{\d h_f}{h_f} + \frac{\d m_f}{m_f} \bigg)
\\[2mm]
&& - g_Z\, m_Z \,e_f \,s_{W}^2\, c_{\a+\b} \bigg( \frac{\d g_Z}{g_Z} + \frac{\d
m_Z}{m_Z} + \frac{\d s_W^2}{s_W^2} - t_{\a+\b}\, \d\b \bigg)
\end{eqnarray}
for the couplings to $H^0$.
\\ %
Analogously to the decays of the CP-even Higgs bosons, the sum of the
wave-function and counter-term corrections of the charged Higgs boson
can be expressed as
\begin{eqnarray}
\d G_{ij1}^{\uparrow\downarrow (w+c)} & = & \d 
G_{ij1}^{\uparrow\downarrow (w,\,\rm symm.)} + \Big[ R^{\sf^\uparrow} 
\cdot \d G_{LR,1}^{\uparrow\downarrow} \cdot ( R^{\sf^\downarrow})^T 
\Big]_{ij} + \d G_{ij1}^{\uparrow\downarrow (w, HW+HG)} 
\end{eqnarray}
with the symmetrized wave-function corrections
\begin{eqnarray}
\d G_{ij1}^{\uparrow\downarrow (w,\,\rm symm.)} = \frac{1}{4} \big(\d 
Z^{\sf^\uparrow}_{ii'} + \d Z^{\sf^\uparrow}_{i'i}\big) 
G_{i'j1}^{\uparrow\downarrow} + \frac{1}{4} \big(\d 
Z^{\sf^\downarrow}_{jj'} + \d Z^{\sf^\downarrow}_{j'j}\big) 
G_{ij'1}^{\uparrow\downarrow} + {\frac{1}{2}}\d Z_{11}^{H^+} 
G_{ij1}^{\uparrow\downarrow}\,. 
\end{eqnarray}
The single elements of the matrix corresponding to the variation with
respect to the couplings, $\d G_{LR,1}^{\uparrow\downarrow}$, are given explicitly 
as follows: 
\begin{eqnarray}\non
\big( \d G_{LR,1}^{\uparrow\downarrow} \big)_{11} & = & 
h_{f_\downarrow} m_{f_\downarrow} s_\b \bigg( \frac{\d 
h_{f_\downarrow}}{h_{f_\downarrow}} + \frac{\d 
m_{f_\downarrow}}{m_{f_\downarrow}} + \frac{\d s_\b}{s_\b} \bigg) + 
h_{f_\uparrow} m_{f_\uparrow} c_\b \bigg( \frac{\d 
h_{f_\uparrow}}{h_{f_\uparrow}} + \frac{\d 
m_{f_\uparrow}}{m_{f_\uparrow}} + \frac{\d c_\b}{c_\b} \bigg) 
\\[2mm]
&& -\frac{g m_{W}}{\sqrt 2} \sin 2\b \bigg( \frac{\d g}{g} + \frac{\d
m_W}{m_W} + \cos 2\b \,\frac{\d\tan\b}{\tan\b} \bigg)
\\[2mm]
\big( \d G_{LR,1}^{\uparrow\downarrow} \big)_{12} & = & \frac{\d 
h_{f_\downarrow}}{h_{f_\downarrow}} \big(G_{LR,1}^{\uparrow\downarrow} 
\big)_{12} + h_{f_\downarrow} \big(\d A_{f_\downarrow} s_\b + 
A_{f_\downarrow} \d s_\b + \d\mu\, c_\b + \mu \,\d c_\b\big) 
\\[2mm]
\big( \d G_{LR,1}^{\uparrow\downarrow} \big)_{21} & = & \frac{\d 
h_{f_\uparrow}}{h_{f_\uparrow}} \big(G_{LR,1}^{\uparrow\downarrow} 
\big)_{21} + h_{f_\uparrow} \big(\d A_{f_\uparrow} c_\b + 
A_{f_\uparrow} \d c_\b + \d\mu\, s_\b + \mu\, \d s_\b\big) 
\\[2mm]\label{eqlist2}
\big( \d G_{LR,1}^{\uparrow\downarrow} \big)_{22} & = &  h_{f_\uparrow} 
m_{f_\downarrow} c_\b \bigg( \frac{\d h_{f_\uparrow}}{h_{f_\uparrow}} + 
\frac{\d m_{f_\downarrow}}{m_{f_\downarrow}} + \frac{\d c_\b}{c_\b} 
\bigg) + h_{f_\downarrow} m_{f_\uparrow} s_\b \bigg( \frac{\d 
h_{f_\downarrow}}{h_{f_\downarrow}} + \frac{\d 
m_{f_\uparrow}}{m_{f_\uparrow}} + \frac{\d s_\b}{s_\b} \bigg) 
\end{eqnarray}
The counter terms appearing in eqs.~(\ref{eqlist1}-\ref{eqlist2}) can 
be fixed in the following manner. Some of them can be decomposed 
further as is the case for $\d h_f$ and $\d g$ 
\begin{eqnarray}
   \frac{\d h_f}{h_f} &=& \frac{\d g}{g} \,+\, \frac{\d m_f}{m_f}
   \,-\, \frac{\d m_{W}}{m_{W}}
   \,+\, {\left\{ \hspace{-5pt}
   \begin{array}{r}{-\cos^2\beta} \\ {\sin^2\beta}
   \end{array} \hspace{-3pt} \right\}}            
   \frac{\d\tan\beta}{\tan\beta}\,,
\quad\qquad\frac{\d g}{g} = \frac{\d e}{e} - \frac{\d\sin\theta_W}{\sin\theta_W} \,,\quad
\end{eqnarray}
for {\scriptsize{$\left\{\hspace{-4pt}\begin{array}{cc}{\textrm{up}}\\ 
{\textrm{down}}\end{array}\hspace{-4pt}\right\}$}}-type sfermions. 
\newline %
For the remaining counter terms we use the standard renormalization 
conditions. The fixing of the angle $\b$ is performed using the 
condition that the renormalized $A^0$-$Z^0$ transition vanishes at $p^2 
= m_{A^0}^2$ as in \cite{pokorski}, which gives the counter term 
\begin{eqnarray}\label{dtanb}
   \frac{\d \tan\b}{\tan\b} &=& \frac{1}{m_{Z}
   \sin 2\b}\, {\rm Im}
   \Pi_{A^0 Z^0} (m_{A^0}^2).
\end{eqnarray}
The higgsino mass parameter $\mu$ is fixed in the chargino sector by 
the chargino mass matrix, $\d\mu \equiv \d X_{22}$, as explained in 
detail in \cite{mass_matrix_corr, Willi}. 
\newline %
The counter term to the Standard Model parameter $\sin\theta_W$ is 
determined using the on-shell masses of the gauge bosons as in 
\cite{sirlin}. To avoid the problems with light quarks in the fine 
structure constant $\a$, we use the $\overline{\rm MS}$ value at the 
$Z$-pole with the counter term given in \cite{A0sq_letter, 
eentntWilli}. 
\newline %
The on-shell counter term that has the biggest influence and also poses 
a serious problem is the counter term to the trilinear scalar coupling 
parameter $A_f$. The explicit form of the counter term was already 
given in \cite{A0sq_letter} and it was shown in \cite{A0sq_letter, 
A0paper} that this counter term becomes very large for large values of 
$\tan\beta$. One of the aims of this paper is to show that this problem 
is present in all Higgs decays into sfermions. The solution takes 
advantage of the fact that the SPA convention which we use here, 
defines the SUSY parameters in the \drbar scheme. Therefore, the 
trilinear scalar coupling parameters $A_f$ are taken \drbar without the 
use of the large on-shell counter term.  \vspace{2mm} 
\section{Infrared divergences}
\begin{figure}[htbp]
\begin{picture}(165,42)(0,0)
     \put(7,0){\mbox{\resizebox{15cm}{!}
     {\includegraphics{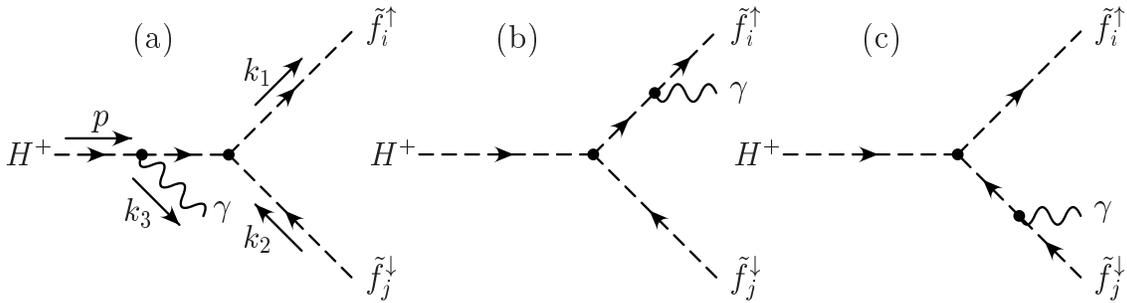}}}}
\end{picture}
\caption{Real Bremsstrahlung diagrams relevant to cancel the 
IR-divergences in $H^+(p) \rightarrow \sf^\uparrow_i(k_1)\, 
+\bar{\sf^\downarrow_j}(-k_2) +\g(k_3)$. The diagrams for the neutral 
Higgs decays are analogous.\label{Hp-radiation}} 
\end{figure}

To cancel infrared divergences we introduce a small photon mass $\l$ 
and include the real photon emission processes $h_k^0 \rightarrow 
\tilde{f}_i\,\,\,{\bar{\!\!\tilde{f}}}_{\!j} \g$ ($h_k^0 = \{h^0, H^0\} 
$) and $H^+ \rightarrow \sf^\uparrow_i \bar{\sf^\downarrow_j} \g$. The 
decay width of $H^+(p) \rightarrow \sf^\uparrow_i(k_1)\, 
+\bar{\sf^\downarrow_j}(-k_2) +\g(k_3)$ (Fig.~\ref{Hp-radiation}) is 
given by 
\begin{eqnarray}\non
&&\G(H^+ \rightarrow \sf^\uparrow_i\, \bar{\sf^\downarrow_j} \g) = \frac{N_C}{16 \pi^3 
m_{H^+}} |G^{\uparrow\downarrow}_{ij1}|^2 (-e_0)^2 \Big[ m_{H^+}^2 I_{00} + e_t^2 
m_i^2 I_{11} + e_b^2 m_j^2 I_{22} 
\\[2mm]\non
&& - e_t e_b \Big( (m_{H^+}^2 - m_i^2 - m_j^2) I_{12} - I_2 - I_1 
\Big) - e_t \Big( (m_j^2 - m_{H^+}^2 - m_i^2) I_{01} - I_1 - I_0 \Big) 
\\[2mm]\non
&&\hphantom{\G(H^+ \rightarrow \st_i \,\bar\sb_j \g) = \frac{N_C}{16 
\pi^3 m_{H^+}} |G^{\uparrow\downarrow}_{ij1}|^2 (-e_0)^2 \Big[} + e_b 
\Big( (m_i^2 - m_{H^+}^2 - m_j^2) I_{02} - I_2 - I_0 \Big) \Big]\,, 
\end{eqnarray}
with the phase-space integrals $I_n$ and $I_{mn}$ defined as 
\cite{Denner} 
\begin{eqnarray}
I_{i_1\ldots i_n}=\frac{1}{\pi^2}
\int\frac{d^3k_1}{2E_1}\frac{d^3k_2}{2E_2}\frac{d^3k_3}{2E_3}
\,\delta^4(p-k_1-k_2-k_3)\frac{1}
{(2k_3k_{i_1}+\lambda^2)\ldots(2k_3k_{i_n}+\lambda^2)} \,.
\end{eqnarray}
The full IR-finite one-loop corrected decay width for the physical 
value $\l = 0$ is then given by 
\begin{eqnarray}\label{correctedwidth}
\G^{\rm corr} (H^+ \rightarrow \sf^\uparrow_i\, \bar{\sf^\downarrow_j}) 
&\equiv& \G(H^+ \rightarrow \sf^\uparrow_i\, \bar{\sf^\downarrow_j}) + 
\G(H^+ \rightarrow \sf^\uparrow_i\, \bar{\sf^\downarrow_j} \g) 
\end{eqnarray}
Analogously, for the neutral Higgs boson decays and the crossed 
channels the photon emission processes yield 
\begin{eqnarray}\non
\G (h_k^0 \rightarrow \tilde{f}_i \ {\bar{\!\!\tilde{f}}}_{\!j}\,\g) 
\!&=&\! N_C\frac{(e_0\,e_f)^2\,|G^{\sf}_{ijk}|^2}{16 \pi^3 m_{h_k^0}} 
\bigg[ \Big(m_{h_k^0}^2 \!-\! m_{\sf_i}^2 \!-\! m_{\sf_j}^2\Big) I_{12} 
+ m_{\sf_i}^2 I_{11} + m_{\sf_j}^2 I_{22} \!-\! I_1 \!-\! I_2 \bigg] 
\,, 
\\[2mm]\non
\G(\sf_2 \rightarrow \sf_1 h_k^0\,\g) \!&=&\! 
\frac{(e_0\,e_f)^2\,|G_{ijk}^{\sf}|^2}{16\,\pi^3\,m_{\sf_2}} 
\bigg[ \Big( m_{h_k^0}^2\!-\!m_{\sf_1}^2\!-\!m_{\sf_2}^2 \Big) I_{01} - 
m_{\sf_1}^2 I_{11} - m_{\sf_2}^2 I_{00} - I_0 - I_1\bigg] \,,
\\ 
\end{eqnarray}
where the IR-finite decay widths are
\begin{eqnarray} 
\G^{\rm corr}(h_k^0 \rightarrow \tilde{f}_i \
{\bar{\!\!\tilde{f}}}_{\!j}) &\equiv& \G(h_k^0 \rightarrow \tilde{f}_i 
\ {\bar{\!\!\tilde{f}}}_{\!j}) \,+\, \G(h_k^0 \rightarrow \tilde{f}_i \ 
{\bar{\!\!\tilde{f}}}_{\!j}\,\g)\,, 
\\[2mm]
\G^{\rm corr}(\sf_2 \rightarrow \sf_1 h_k^0) &\equiv& \G(\sf_2 
\rightarrow \sf_1 h_k^0) \,+\, \G(\sf_2 \rightarrow \sf_1 h_k^0\,\g)\,.
\end{eqnarray}
\section{Numerical analysis}
\begin{figure}[h!]
\begin{picture}(160,68)(0,0)
    \put(10,65){$\{ M_{{\ti D}_1} = 150 \GeV,\ m_{A^0}^2 = 10^6\GeV \}$}
    \put(30,5){\mbox{\resizebox{8.6cm}{!}
    {\includegraphics{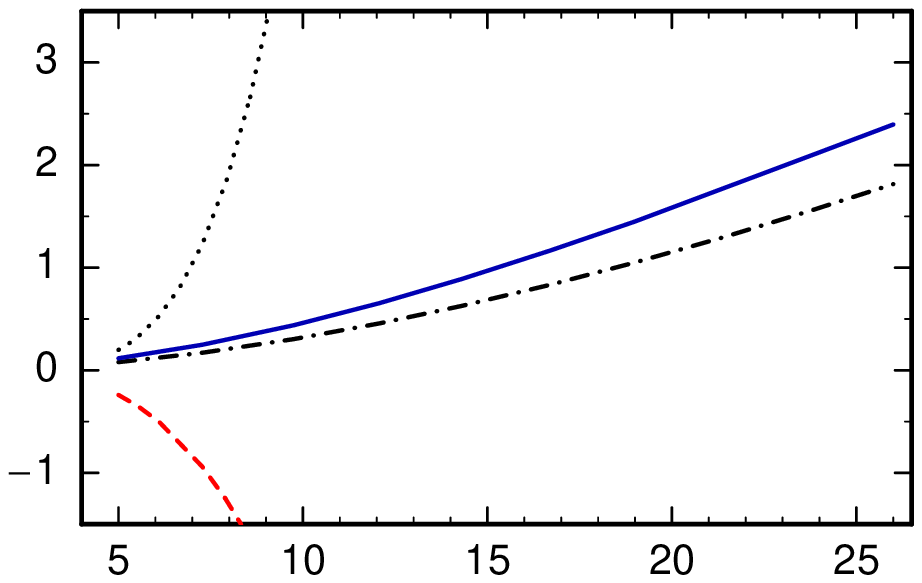}}}}
    \put(78,3){\makebox(0,0)[t]{{$\tan\b$}}}
    \put(23,15){\rotatebox{90}{{{$\Gamma (H^0 \rightarrow
                        \sd_1 \!\bar{\,\sd_2})$\ [MeV]}}}}
\end{picture}\caption{On-shell tree-level (dotted line), full on-shell one-loop 
decay width (dashed line), tree-level (dash-dotted line) and full 
one-loop corrected width (solid line) of $H^0 \rightarrow \sd_1 
\!\bar{\,\sd_2}$ as a function of $\tan\b$ according to the SPA 
convention. \label{H0sd1sd2_tanb}} 
\end{figure}

\begin{figure}[h!]
\begin{picture}(160,60)(0,0)
    \put(30,5){\mbox{\resizebox{8.6cm}{!}
    {\includegraphics{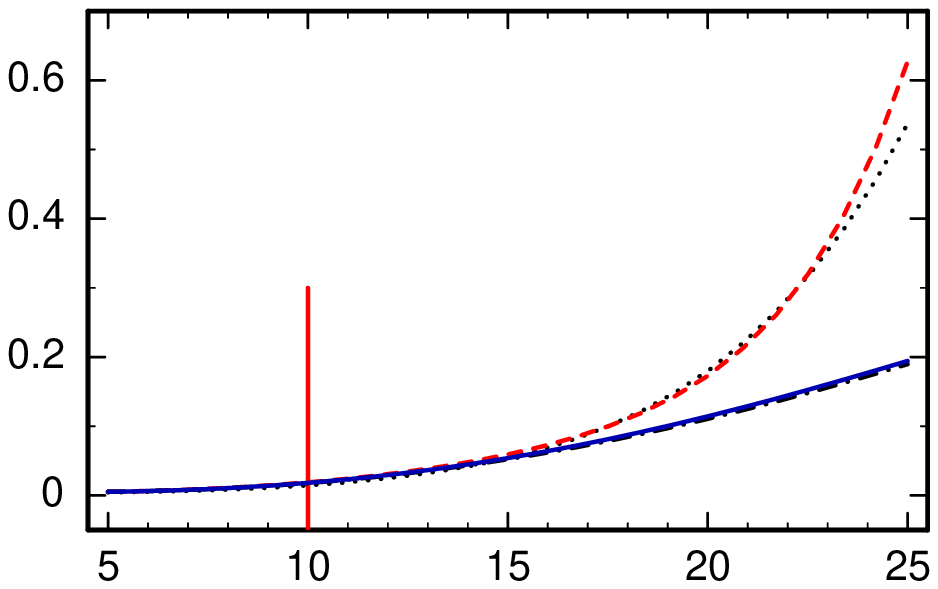}}}}
    \put(78,3){\makebox(0,0)[t]{{$\tan\beta$}}}
    \put(23,15){\rotatebox{90}{{{$\Gamma (H^0 \rightarrow
                        \stau_1\stau_1)$\ [GeV]}}}}
\end{picture}\caption{On-shell tree-level (dotted line), full on-shell one-loop 
decay width (dashed line), tree-level (dash-dotted line) and full 
one-loop corrected width (solid line) of $H^0 \rightarrow 
\stau_1\stau_1$ as a function of $\tan\beta$ according to the SPA 
convention. \label{H0sl1sl1_tanb}} 
\end{figure}

\begin{figure}[h!]
\begin{picture}(160,68)(0,0)
    \put(10,65){$\{ M_{{\ti D}_3} = 150 \GeV,\ m_{A^0}^2 = 10^6\GeV \}$}
    \put(30,5){\mbox{\resizebox{8.6cm}{!}
    {\includegraphics{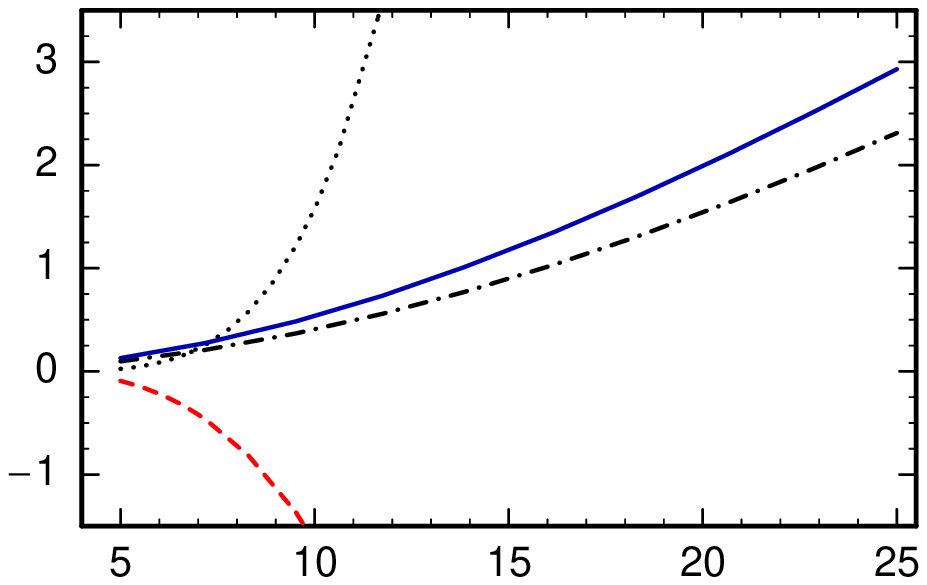}}}}
    \put(77,3){\makebox(0,0)[t]{{$\tan\beta$}}}
    \put(23,15){\rotatebox{90}{{{$\Gamma (H^+ \rightarrow
                        \st_1 \!\bar{\,\sb_2})$\ [GeV]}}}}
\end{picture}\caption{On-shell tree-level (dotted line), full on-shell 
one-loop decay width (dashed line), tree-level (dash-dotted line) and 
full one-loop corrected width (solid line) of $H^+ \rightarrow \st_1 
\!\bar{\,\sb_2}$ as a function of $\tan\beta$ according to the SPA 
convention. \label{Hpst1sb2_tanb}} 
\end{figure}

\begin{figure}[h!]
\begin{picture}(160,68)(0,0)
    \put(10,65){$\{ M_{{\ti D}_2} = 150 \GeV,\ m_{A^0}^2 = 10^6\GeV \}$}
    \put(30,5){\mbox{\resizebox{8.6cm}{!}
    {\includegraphics{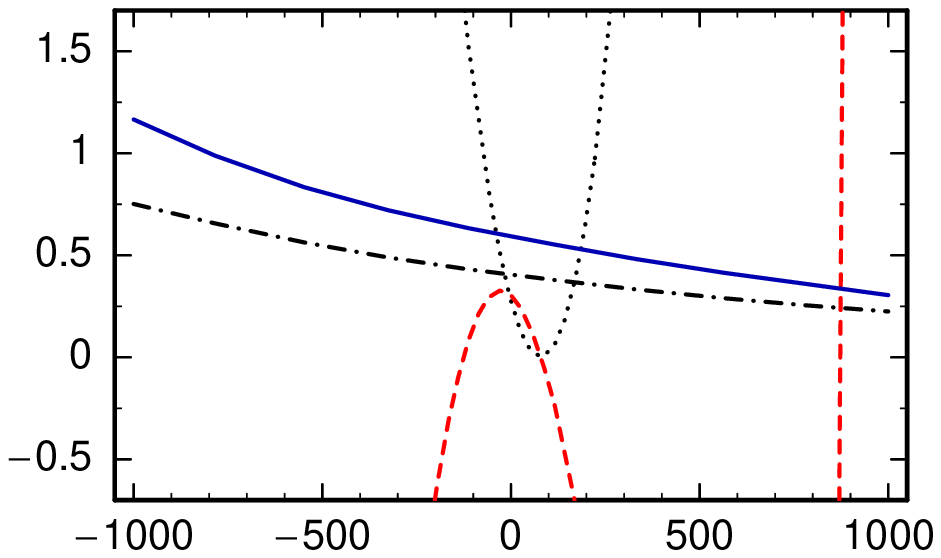}}}}
    \put(77,3){\makebox(0,0)[t]{{$\mu$\ [GeV]}}}
    \put(23,15){\rotatebox{90}{{{$\Gamma (H^0 \rightarrow
                        \sstrange_1 \!\bar{\,\sstrange_2})$\ [MeV]}}}}
\end{picture}\caption{On-shell tree-level (dotted line), full on-shell one-loop 
decay width (dashed line), tree-level (dash-dotted line) and full 
one-loop corrected width (solid line) of $H^0 \rightarrow \sstrange_1 
\!\bar{\,\sstrange_2}$ as a function of $\mu$ according to the SPA 
convention. \label{H0ss1ss2_xmu}} 
\end{figure}

\begin{figure}[h!]
\begin{picture}(160,68)(0,0)
    \put(10,65){$\{ M_{{\ti U}_3} = 150 \GeV,\ m_{A^0}^2 = 10^6\GeV \}$}
    \put(30,5){\mbox{\resizebox{8.6cm}{!}
    {\includegraphics{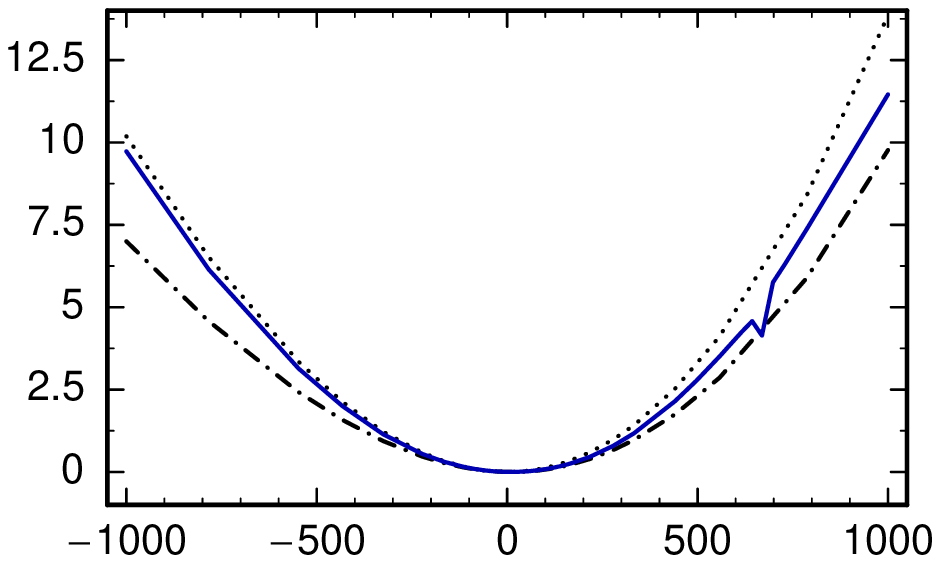}}}}
    \put(77,3){\makebox(0,0)[t]{{$\mu$\ [GeV]}}}
    \put(23,15){\rotatebox{90}{{{$\Gamma (H^0 \rightarrow
                        \st_1 \!\bar{\,\st_1})$\ [GeV]}}}}
\end{picture}\caption{On-shell tree-level (dotted line), full on-shell one-loop 
decay width (dashed line), tree-level (dash-dotted line) and full 
one-loop corrected width (solid line) of $H^0 \rightarrow \st_1 
\!\bar{\,\st_2}$ as a function of $\mu$ according to the SPA 
convention. \label{H0st1st1_xmu}} 
\end{figure}

\begin{figure}[h!]
\begin{picture}(160,60)(0,0)
    \put(30,5){\mbox{\resizebox{8.6cm}{!}
    {\includegraphics{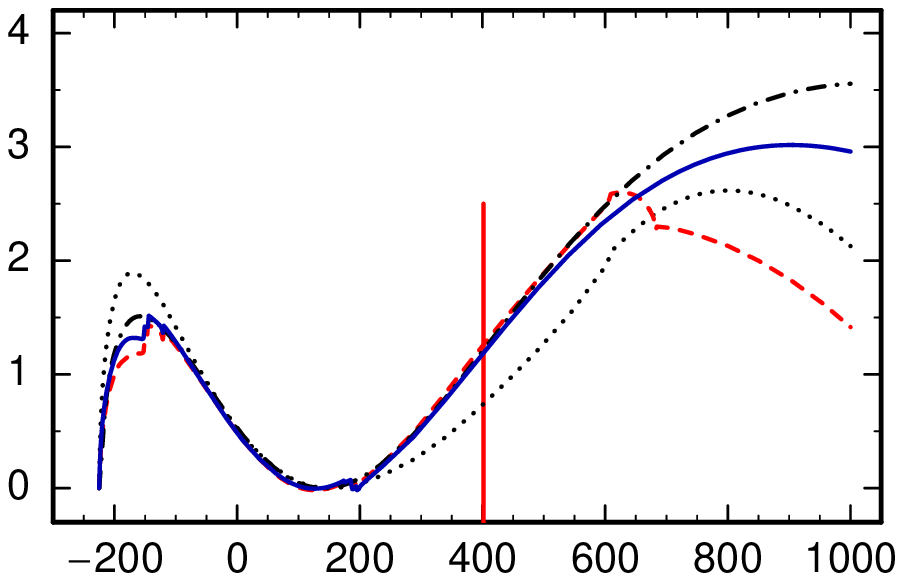}}}}
    \put(77,3){\makebox(0,0)[t]{{$\mu$\ [GeV]}}}
    \put(23,15){\rotatebox{90}{{{$\Gamma (H^+ \rightarrow
                        \snutau \!\bar{\,\stau_2})$\ [GeV]}}}}
\end{picture}\caption{On-shell tree-level (dotted line), full on-shell one-loop 
decay width (dashed line), tree-level (dash-dotted line) and full 
one-loop corrected width (solid line) of $H^+ \rightarrow \snutau 
\!\bar{\,\stau_2}$ as a function of $\mu$ according to the SPA 
convention. \label{Hpslnsl2_xmu}} 
\end{figure}

The numerical results presented in this section are based on the SPS1a' 
benchmark point as proposed by the Supersymmetric Parameter Analysis 
Project (SPA) \cite{SPA}. A consistent implementation of the SPA 
convention into the calculation of a decay width and the numerical 
analysis is a non-trivial endeavor. As the electroweak one-loop 
calculations are carried out in the on-shell scheme and the SPA project 
proposes the SUSY input set in the \drbar scheme at the scale of 
$1\TeV$, a conversion of the input values is necessary. This conversion 
requires the renormalization of the whole MSSM in order to transform 
the input parameters correctly. Moreover, the numerical analysis of a 
decay makes varying fundamental SUSY parameters necessary. That is why 
the above mentioned transformation of parameters has to be carried out 
for every single parameter point. In our case this is provided by the 
not-yet-public routine DRbar2OS which couples to the spectrum 
calculator SPheno \cite{SPheno}. The transformation is performed in the 
following two steps: 
\begin{enumerate}
\item The SPA input, i.e. the on-shell electroweak SM parameters, the 
strong coupling constant and the masses of the light quarks defined in 
the \msbar scheme and the masses of the leptons and the top quark 
defined as pole masses and the SUSY parameters defined in the \drbar 
scheme at 1\TeV, is given to SPheno which transforms it to a pure 
\drbar input set including also higher loop corrections. 
\item The pure \drbar set is taken as input for the DRbar2OS routine which yields 
as output the complete set in the on-shell scheme. An example of different sets of parameters 
for the SPS1a' benchmark point can be seen in Table~\ref{SPAinp}. 
\end{enumerate}
All plots below show the dependence of the decay width on a \drbar 
parameter. By varying a single \drbar parameter and transforming 
subsequently to the on-shell scheme, almost all parameters are 
influenced. That means, not only the corresponding on-shell parameter 
changes, but also the other parameters through loop effects. 
\begin{table}[h!]
\begin{center}
\renewcommand{\arraystretch}{1.5}
{\small 
\begin{tabular}{|c|c||c|c|}
\hline \multicolumn{4}{|c|}{\drbar SUSY Parameters}\\ \hline 
 $g'$ &  0.36354  &  $M_1$        & 103.21 \\
 $g$ &   0.64804  &  $M_2$        & 193.29 \\                         
 $g_s$ & 1.08412  &  $M_3$        & 572.33 \\ \hline
 $Y_\tau$ & 0.10349 & $A_\tau $     & $-445.4$\\
 $Y_t$ &    0.89840 & $A_t $        & $-532.3$ \\
 $Y_b$ &    0.13548 & $A_b $        & $-938.9$  \\ \hline
$\mu$  & 401.63   & $\tan\beta$   & 10.0    \\ \hline 
 $M_{L_1}$  &$1.8121 \cdot 10^2$ &  $M_{L_3}$  &$1.7945 \cdot 10^2$  \\
 $M_{E_1}$  &$1.1572 \cdot 10^2$ &  $M_{E_3}$  &$1.1002 \cdot 10^2$  \\
 $M_{Q_1}$  &$5.2628 \cdot 10^2$ &  $M_{Q_3}$  &$4.7091 \cdot 10^2$\\
 $M_{U_1}$  &$5.0767 \cdot 10^2$ &  $M_{U_3}$  &$3.8532 \cdot 10^2$\\
  $M_{D_1}$ &$5.0549 \cdot 10^2$ &  $M_{D_3}$  &$5.0137 \cdot 10^2$\\
\hline $M^2_{H_1} $  &$2.5605 \cdot 10^4$  & $M^2_{H_2} $ & $-14.725 
\cdot 10^4$\\ \hline 
\end{tabular}
\quad 
\begin{tabular}{|c|c||c|c|}
\hline \multicolumn{4}{|c|}{On-shell SUSY Parameters}\\ \hline 
 $g'$ &  0.35565  &  $M_1$        & 100.31 \\
 $g$ &   0.66547  &  $M_2$        & 197.01 \\                         
 $g_s$ & 1.08419  &  $M_3$        & 612.81 \\ \hline
 $Y_\tau$ & 0.10771 & $A_\tau $     & $-394.2$\\
 $Y_t$ &    1.04638 & $A_t $        & $-495.0$ \\
 $Y_b$ &    0.20481 & $A_b $        & $1197.8$  \\ \hline
$\mu$  & 398.71   & $\tan\beta$   & 10.31    \\ \hline 
 $M_{L_1}$  &$1.8394 \cdot 10^2$ &  $M_{L_3}$  &$1.8199 \cdot 10^2$  \\
 $M_{E_1}$  &$1.1784 \cdot 10^2$ &  $M_{E_3}$  &$1.1172 \cdot 10^2$  \\
 $M_{Q_1}$  &$5.6390 \cdot 10^2$ &  $M_{Q_3}$  &$5.0369 \cdot 10^2$\\
 $M_{U_1}$  &$5.4540 \cdot 10^2$ &  $M_{U_3}$  &$4.1021 \cdot 10^2$\\
  $M_{D_1}$ &$5.4352 \cdot 10^2$ &  $M_{D_3}$  &$5.3894 \cdot 10^2$\\
\hline $M^2_{H_1} $  &$2.7220 \cdot 10^4$  & $M^2_{H_2} $ & $-15.726 
\cdot 10^4$\\ \hline 
\end{tabular}
} 
\renewcommand{\arraystretch}{1}
\end{center}
\caption{The input parameters for the SPS1a' point according to the SPA 
project. \hspace{1cm}Left: \drbar input values at $Q=1\TeV$, Right: 
on-shell values used in the calculation.}\label{SPAinp} 
\end{table}
\newline %
Comparing the parameter sets in Table \ref{SPAinp} one can easily see 
that the on-shell counter term for $A_b$ is large as there is a huge 
difference between the \drbar and the on-shell value of the trilinear 
scalar coupling parameter. This is caused by fixing the of $A_b$ 
parameter in the sfermion sector \cite{A0sq_letter}. This fixing was 
shown to lead to a numerically large counter term which should be 
avoided. The decays of Higgs bosons into sfermions (and the 
corresponding crossed channels) are the only 2-body decays that are 
affected directly as the trilinear coupling parameter appears at 
tree-level. In this case, a very large counter term makes the 
perturbative expansion unreliable. Here we make use of the fact that 
the input parameters are given in the \drbar scheme. That means our 
calculation uses on-shell parameters except for the $A_b$ and $m_b$ 
which take the original \drbar values. Although not shown in the Table 
\ref{SPAinp}, this behaviour is common to all down-type trilinear 
scalar coupling parameters, and the same strategy as described for the 
$A_b$ is applied for them as well. 
\newline %
A distinct feature of all decay modes involving down-type sfermions is 
a large difference between the on-shell and the SPA tree-level. The 
origin of this difference is again the large counter term for the 
trilinear scalar couplings.
\newline %
Keeping the numerical analysis strictly confined to the SPS1a' 
benchmark scenario would mean that most of the possible decays are 
kinematically not allowed. The vertical red line denotes the position 
of the SPS1a' parameter point for the kinematically allowed decay 
modes. For the other decays we slightly deviate from the SPS1a' point 
adjusting mainly $m_{A^0}$ and the relevant soft supersymmetry breaking 
terms $M_{\{\ti Q,\ti U,\ti D,\ti L,\ti E\}}$. These parameters only 
influence the kinematics and have usually no effect on the couplings. 
\newline %
In general, we always show the results using the on-shell parameters 
(dotted curve for on-shell tree-level and red dashed curve for on-shell 
full one loop decay width) as well as the improved decay widths where 
the parameters $A_f$ and $m_f$ are taken \drbar (dash-dotted curve 
denotes SPA tree-level and blue solid curve stands for the full one 
loop decay width). This convention does not apply in cases where there 
is no down-type trilinear scalar coupling entering the tree-level. 
There we show only the on-shell and SPA tree-level together with the 
final one-loop decay width. 
For comparison with other calculations, the SPA tree-level is shown as 
defined in \cite{SPA} taking all parameters in the couplings in the 
\drbar scheme and using the proper masses for the kinematics.
\newline\newline %
\begin{figure}[h!]
\begin{picture}(160,68)(0,0)
    \put(10,65){$\{ M_{{\ti D}_3} = 150 \GeV,\ m_{A^0}^2 = 10^6\GeV \}$}
    \put(30,5){\mbox{\resizebox{8.6cm}{!}
    {\includegraphics{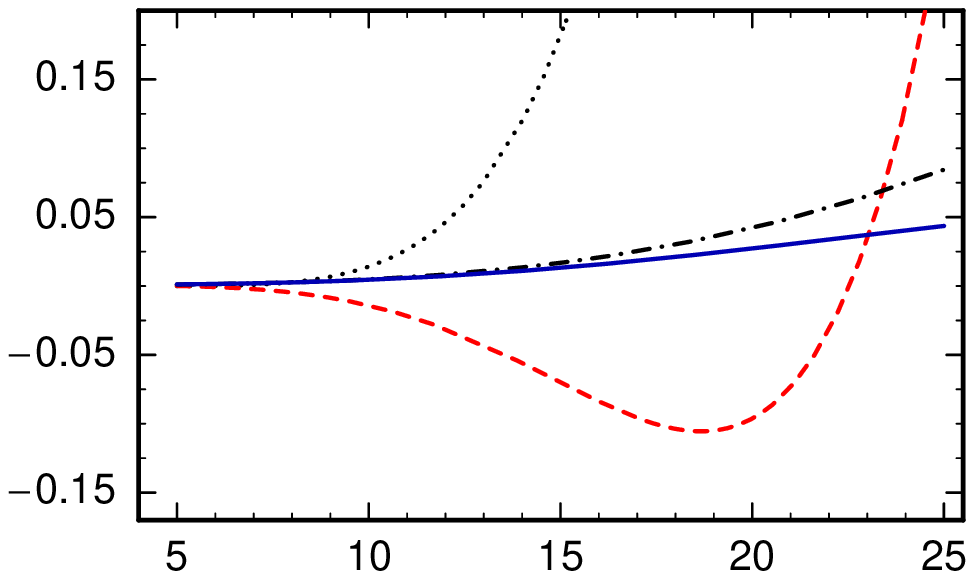}}}}

    \put(80,4){\makebox(0,0)[t]{{$\tan\beta$}}}
    \put(23,15){\rotatebox{90}{{{$\Gamma (H^0 \rightarrow
                        \sb_1 \!\bar{\,\sb_1})$\ [GeV]}}}}
\end{picture}\caption{On-shell tree-level (dotted line), full on-shell one-loop 
decay width (dashed line), tree-level (dash-dotted line) and full 
one-loop corrected width (solid line) of $H^0 \rightarrow \sb_1 
\!\bar{\,\sb_1}$ as a function of $\tan\beta$ according to the SPA 
convention. \label{H0sb1sb1_tanb}} 
\end{figure}
%
\newline %
Figs.~\ref{H0sb1sb1_tanb}, \ref{H0sd1sd2_tanb} and \ref{H0sl1sl1_tanb} 
show the dependence of the decay width on $\tan\beta$ and clearly 
demonstrate the known fact \cite{SUSY-QCD, A0sq_letter} that the counter term for 
$A_b$ grows with $\tan\beta$. As mentioned above such a large counter 
term causes the perturbation series to break down. The full one-loop 
results in the on-shell scheme (red dashed curve) differ from those 
where \drbar parameters are used (blue solid) only by higher orders. 
Nevertheless, due to the perturbation series breakdown, the higher 
order corrections are no longer suppressed by the coupling constant.
This is the reason why the results using the \drbar parameters should 
be viewed as the final one-loop corrected decay widths for the 
processes calculated in this paper. 
%
%
\clearpage %
\noindent The next class of plots shows the dependence on the 
superpotential parameter $\mu$ (see Figs.~\ref{H0ss1ss2_xmu}, 
\ref{H0st1st1_xmu}, \ref{Hpslnsl2_xmu} and \ref{Hpst1sb2_tanb}). The 
typical behaviour of the tree-level is governed by the square of the 
coupling $G^{\sf}_{ijk}$. It implies that the tree-level is a quadratic 
function of $\mu$ and as all one-loop corrections in this case are 
factorizable this dependence is preserved in the full one-loop result. 
In case down-type sfermions are involved, the on-shell curves are 
deformed by the large difference in the $A_f$ parameter. The 
corrections can reach up to 40\% for some areas of parameter space and 
are therefore not negligible. 
%

\begin{figure}[h!]
\begin{picture}(160,68)(0,0)
    \put(10,65){$\{ M_{{\ti D}_3} = 150 \GeV \}$}
    \put(30,5){\mbox{\resizebox{8.6cm}{!}
    {\includegraphics{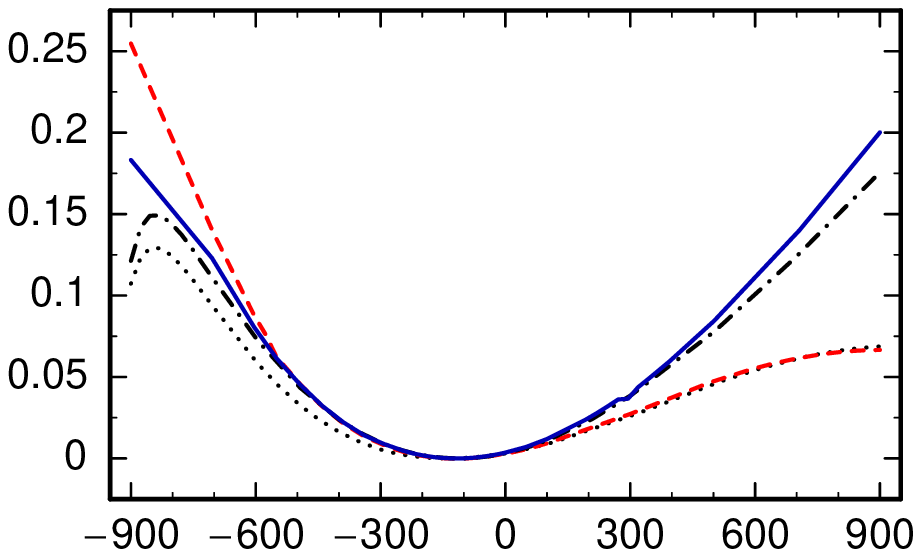}}}}
    \put(77,3){\makebox(0,0)[t]{{$\mu$\ [GeV]}}}
    \put(23,15){\rotatebox{90}{{{$\Gamma (\sb_2 \rightarrow
                        \sb_1 h^0)$\ [GeV]}}}}
\end{picture}\caption{On-shell tree-level (dotted line), full on-shell one-loop 
decay width (dashed line), tree-level (dash-dotted line) and full 
one-loop corrected width (solid line) of $\sb_2 \rightarrow \sb_1 h^0$ 
as a function of $\mu$ according to the SPA convention. 
\label{sb2h0sb1_xmu}} 
\end{figure}

%
\clearpage\noindent Furthermore, the pseudothreshold in 
Fig.~\ref{H0st1st1_xmu} comes from the sbottom contribution to the 
Higgs wave-function correction. 
\newline %
The one-loop width of $H^+ \rightarrow \snutau \!\bar{\,\stau_2}$ (see 
Fig.~\ref{Hpslnsl2_xmu}) is unexpectedly sensitive to the large 
difference of the on-shell and \drbar $A_b$ parameters above $\mu=600 
\GeV$ although there is no such parameter at tree-level. It is caused 
by the enhanced contribution of the vertex diagram with a 4-sfermion 
coupling. This diagram contains the coupling of the charged Higgs boson 
and a stop-sbottom pair where the $A_b$ parameter appears. 
%
%
\clearpage\noindent Fig.~\ref{sb2h0sb1_xmu} illustrates the 
aforementioned problems of the perturbation series in case of using the 
on-shell $A_b$ parameter. As one can see there is no obvious divergence 
of the decay width in the on-shell scheme. Nevertheless, the full 
one-loop widths in the on-shell scheme and the one with $A_b$ taken 
\drbar are far apart. This separation is a pure two loop effect coming 
from using different $A_b$ values when calculating the $\d A_b$ counter 
term. 
\newline %
In this particular case the electroweak corrections interfere 
destructively with the QCD corrections reducing them by half. 
%
\begin{figure}[h!]
\begin{picture}(160,60)(0,0)
    \put(30,5){\mbox{\resizebox{8.6cm}{!}
    {\includegraphics{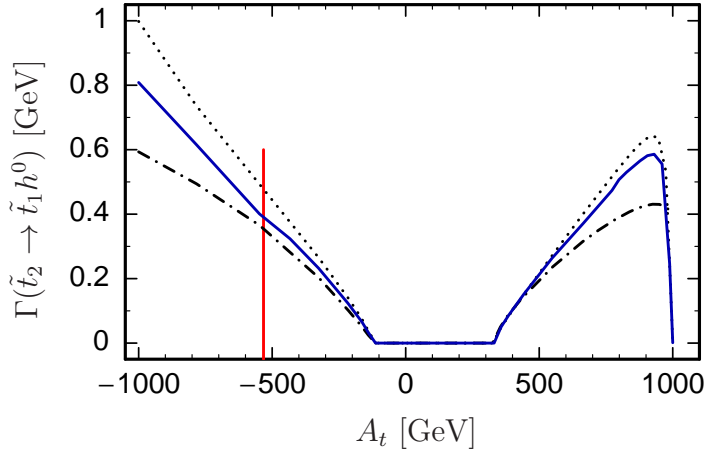}}}}
    \put(77,3){\makebox(0,0)[t]{{$A_t$\ [GeV]}}}
    \put(23,16){\rotatebox{90}{{{$\Gamma (\st_2 \rightarrow
                        \st_1 h^0)$\ [GeV]}}}}
\end{picture}\caption{On-shell tree-level (dotted line), 
tree-level (dash-dotted line) and full one-loop corrected width (solid 
line) of $\st_2 \rightarrow \st_1 h^0$ as a function of $A_t$ according 
to the SPA convention. \label{st2h0st1_At}} 
\end{figure}
\newline\newline %
The only plot over the trilinear coupling $A_f$ is shown in 
Fig.~\ref{st2h0st1_At} which is for the decay $\st_2 \rightarrow \st_1 
h^0$. The region $A_t = (-120\GeV, 320\GeV)$ is kinematically 
forbidden. Although at some regions of parameter space (e.g. $A_t = 
(700\GeV, 900\GeV)$) the correction to the SPA tree-level is large, one 
can see there is also a large difference between SPA tree-level and the 
tree-level used in the perturbation expansion denoted by the dotted 
line. 
\section{Conclusions}\label{conclusions}
\vspace{2mm} We have calculated the {\em full} electroweak one-loop 
corrections to the charged and CP-even neutral Higgs boson decays to 
sfermions including the crossed channels. We have also included the 
SUSY-QCD corrections which were calculated in \cite{SUSY-QCD}. Similar 
to \cite{A0sq_letter} and \cite{A0paper}, the on-shell parameters $A_b$ 
and $m_b$ (and the corresponding down-type parameters for the first two 
generations) were replaced by their \drbar values to avoid the 
numerically large counter term. Furthermore, we have presented the 
first consistent numerical analysis for a one-loop decay width based on 
the Supersymmetric Parameter Analysis project \cite{SPA}. This required 
the renormalization of the whole MSSM in a way that allows to carry out 
a transformation between the on-shell and the \drbar scheme for every 
single parameter point. 
\newline %
The corrections were shown not to be negligible and were comparable in 
the magnitude to the QCD in some regions of parameter space. 
\newline %
\newline %
\noindent {\bf Acknowledgements}\\ \noindent We thank W.~Porod for his 
generous support in including SPheno in our numerical calculations and 
for useful discussions. The authors acknowledge support from EU under 
the MRTN-CT-2006-035505 network programme and from the "Acciones 
Integradas 2005-2006", project No. 13/2005. This work is supported by 
the "Fonds zur F\"orderung der wissenschaftlichen Forschung" of 
Austria, project No. P18959-N16. 

\clearpage 
\appendix 
\section{Self-energies and counter terms}\label{selfenergies} 
Here we give the explicit form of the self-energies needed for the 
computation of the one-loop decay widths $\{ h^0, H^0 \} \rightarrow 
\sf_i\, \bar{\!\sf_j}$ and $H^+ \rightarrow \sf^\uparrow_i 
\bar{\sf^{\downarrow}_{j}}$. 
\newline%
For the neutral and charged Higgs fields we use the notation $h_k^0 = 
\{h^0, H^0\}$, $H_k^0 = \{h^0, H^0, A^0, G^0\}$, $H_k^+ = \{H^+, G^+, 
H^-, G^-\}$ and $H_k^{-} \equiv (H_k^+)^\ast = \{H^-, G^-, H^+, G^+\}$. 
\newline %
The sfermion self-energies and their derivatives as well as the vector 
boson self-energies and the scalar-vector mixing of $A^0 Z^0$ have 
already been published and can be found in \cite{A0paper}. Also most of 
the couplings used in this paper are given therein. 
\newline
In the following we use the standard two- and three-point functions 
$B_i$ and $C_i$ from \cite{PaVe} in the conventions of \cite{Denner}. 
\subsection{Diagonal Wave-function corrections --- derivatives of 
Higgs boson self-energies} %
\begin{figure}[thbp]
\begin{picture}(165,70)(0,0)
     \put(20,0){\mbox{\resizebox{13cm}{!}
     {\includegraphics{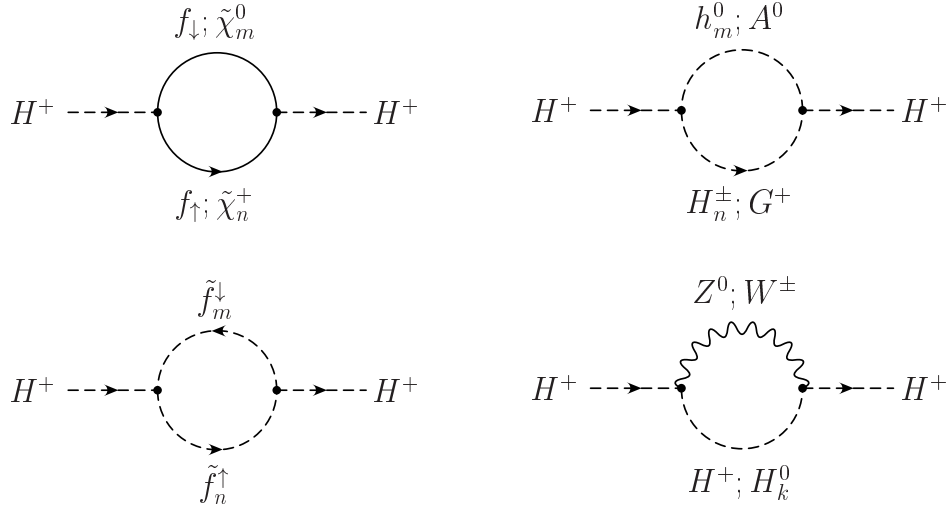}}}}
\end{picture}
\caption{Self-energy diagrams of the charged Higgs boson $H^+$ relevant 
for the calculation of diagonal wave-function corrections. 
\label{HpSEdiag}} 
\end{figure}

The conventional on-shell renormalization conditions for the diagonal 
wave-function renormalization constants are given in terms of the 
derivatives of the corresponding self-energies, 
\begin{eqnarray}
\d Z_{kk}^{H^0} ~=~ - \Re\,\dot\Pi_{kk}^{H^0} (m_{H_k^0}^2) \,, 
\end{eqnarray}
where the dot in $\dot\Pi_{\ldots} (k^2)$ denotes the derivative with 
respect to $k^2$. In the following we list the single contributions of 
the Higgs wave-function corrections. The derivatives of the CP-even 
Higgs bosons $h^0$ and $H^0$ depicted in Fig.~\ref{hk0SEdiag} are given 
as follows: 
\begin{figure}[tbhp]
\begin{picture}(165,105)(0,0)
     \put(0,0){\mbox{\resizebox{16cm}{!}
     {\includegraphics{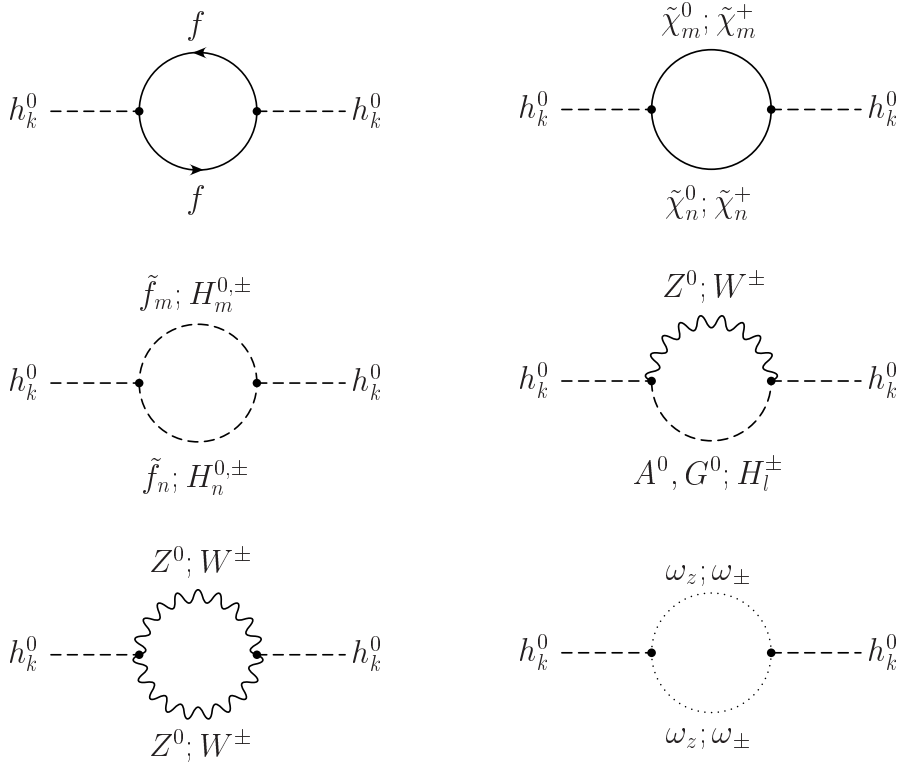}}}}
\end{picture}
\caption{Diagonal self-energies of CP-even Higgs bosons $h^0$ and 
$H^0$\label{hk0SEdiag}} 
\end{figure}
%
\begin{eqnarray}
\dot\Pi_{kk}^{H^0,f} &=& -\frac{2}{(4\pi)^2}\, \sum_{f} N_C^f \big( 
s^f_k \big)^2 \Big[ (4 m_f^2 - m_{h_k^0}^2) \dot B_0( m_{h_k^0}^2, 
m_f^2, m_f^2 ) - B_0( m_{h_k^0}^2, m_f^2, m_f^2 )\Big] 
\\
\dot\Pi_{kk}^{H^0,\sf} &=& \vor\, \sum_{f} \sum_{m,n=1}^2 N_C^f\, 
\big(G^\sf_{mnk}\big)^2\, \dot{B}_0( m_{h_k^0}^2, m_{\sf_m}^2, 
m_{\sf_n}^2 ) 
\\ \non
\dot\Pi_{kk}^{H^0,\nt} &=& -\vor\, g^2 \sum_{m,n=1}^4 (F^0_{mnk})^2 
\Big[ \Big( (m_{\nt_m}+m_{\nt_n})^2 - m_{h_k^0}^2 \Big) \dot{B}_0 - B_0 
\Big] (m_{h_k^0}^2, m_{\nt_m}^2, m_{\nt_n}^2) 
\\
\\ \non
\dot\Pi_{kk}^{H^0,\chp} &=& -\vor\, g^2 \sum_{m,n=1}^2 \Big[ \Big( 
(F^+_{mnk})^2 + (F^+_{nmk})^2 \Big) \Big( \big( m_{\chp_m}^2 + 
m_{\chp_n}^2 - m_{h_k^0}^2 \big) \dot{B}_0 - B_0 \Big) 
\\
&&\hphantom{-\vor\, g^2 \sum_{m,n=1}^2 \Big[} + 4 m_{\chp_m} m_{\chp_n} 
F^+_{mnk} \, F^+_{nmk} \, \dot{B}_0 \Big] (m_{h_k^0}^2, m_{\chp_m}^2, 
m_{\chp_n}^2) 
\\[2mm] \non
\dot\Pi_{kk}^{H^0,H} &=& \vor \frac{1}{2} \left(\frac{g_{Z} \, 
m_{Z}}{4}\right)^2 \bigg[ \sum_{m,n = 1}^2 [(2\!+\!\d_{km} \d_{mn})!]^2 
\big( \cos 2\a \,\ti A_{mn}^{(k)} - 2 \sin 2\a\, \ti B_{mn}^{(k)} 
\big)^2 \dot{B}_0 
\\ \non
&& + 4 \sum_{m,n = 3}^4 \sin^2 [\a+\b-\frac{\pi}{2}(k-1)] \, \big( \ti 
C_{m-2,n-2} \big)^2 \dot{B}_0 \bigg] (m_{h_k^0}^2, m_{H_m^0}^2, 
m_{H_n^0}^2) 
\\ \non
&& +\vor \sum_{m,n = 1}^2 \Big[ (-1)^{mn}\, \frac{g \, m_{W}}{2} 
(1-\d_{m2}\d_{n2}) (1+\d_{mn}) \ti A_{mn}'^{(k)} - \frac{g_{Z} \, 
m_{Z}}{2} 
\\
&&\hphantom{+\vor \sum_{m,n = 1}^2 \bigg[} \times \sin[\a+\b 
-\frac{\pi}{2}(k-1)] \, \ti C_{mn} \Big]^2 \dot{B}_0 (m_{h_k^0}^2, 
m_{H_m^+}^2, m_{H_n^+}^2) 
\\[2mm] \non
\dot\Pi_{kk}^{H^0,V} &=& -\vor\, \frac{g^2}{2} \sum_{l=1}^2 \big( 
R_{lk}(\a\!-\!\b) \big)^2
\\ \non
&&\hspace{3.5cm}\times \Big[ \big( 2 m_{h_k^0}^2 \!+\! 2 m_{H_l^+}^2 
\!-\! m_W^2 \big) \dot{B}_0 + 2 B_0 \Big] (m_{h_k^0}^2, m_{H_l^+}^2, 
m_W^2) 
\\ \non
&& -\vor\, \frac{g_Z^2}{4} \sum_{l=1}^2 \big( R_{lk}(\a\!-\!\b) \big)^2 
\\ \non
&&\hspace{3.5cm}\times \Big[ \big( 2 m_{h_k^0}^2 \!+\! 2 m_{H_{l+2}^0}^2 \!-\! m_Z^2 \big) 
\dot{B}_0 \!+\! 2 B_0 \Big] (m_{h_k^0}^2, m_{H_{l+2}^0}^2, m_Z^2) 
\\
\\ \non
\dot\Pi_{kk}^{H^0,VV} &=& \vor \big( R_{2k}(\a\!-\!\b) \big)^2 \Big[ 4 
g^2 m_W^2 \dot{B}_0 (m_{h_k^0}^2, m_W^2, m_W^2) + 2 g_Z^2 m_Z^2 
\dot{B}_0 (m_{h_k^0}^2, m_Z^2, m_Z^2) \Big] 
\\
\\ \non
\dot\Pi_{kk}^{H^0,{\rm ghost}} &=& -\vor \big( R_{2k}(\a\!-\!\b) 
\big)^2 \Big[ \frac{g^2}{2} m_W^2 \dot{B}_0 (m_{h_k^0}^2, m_W^2, m_W^2) 
+ \frac{g_Z^2}{4} m_Z^2 \dot{B}_0 (m_{h_k^0}^2, m_Z^2, m_Z^2) \Big] \,,
\\
\end{eqnarray}
where we have used 
\begin{eqnarray}\label{Hk0Hl0Hm0_coup_matrices}\non
   \ti A_{mn}^{(k)} &=& \left(
   \begin{array}{r@{\hspace{8mm}}r}
      -\sin[\a+\b-\frac{\pi}{2}(k-1)] & \cos[\a+\b-\frac{\pi}{2}(k-1)]
      \\ 
       \cos[\a+\b-\frac{\pi}{2}(k-1)] & \sin[\a+\b-\frac{\pi}{2}(k-1)]
   \end{array}
   \right)\,,
\\[3mm]\non
   \ti A_{mn}'^{(k)} &=& \ti A_{mn}^{(k)}(\b \rightarrow -\b) \,,
\\[3mm]\non
   \ti B_{mn}^{(k)} &=& \left(
   \begin{array}{cc}
      (k-1) \sin(\a+\b) & \sin[\a+\b-\frac{\pi}{2}(k-1)]
      \\ 
      \sin[\a+\b-\frac{\pi}{2}(k-1)] & (k-2) \cos(\a+\b)
   \end{array}
   \right)\,,
\\[3mm]
   \ti C_{mn} &=& \left(
   \begin{array}{rr}
      \cos 2\b & \sin 2\b
      \\ 
      \sin 2\b & -\cos 2\b
   \end{array}
   \right)\,.
\end{eqnarray}
The diagrams in Fig.~\ref{HpSEdiag} show the diagonal charged Higgs 
boson self-energies entering in the wave-function corrections.
%
\begin{eqnarray}\non
\dot\Pi_{11}^{H^+,f} &=& -\vor \sum_{f=\{f_\uparrow\}} N_C^f \Big[ 
\big( h_{f_\uparrow}^2 \cos^2\b + h_{f_\downarrow}^2 \sin^2\b \big) 
\big( (m_{f_\uparrow}^2 + m_{f_\downarrow}^2 - m_{H^+}^2) \dot B_0 - 
B_0 \big) 
\\
&&\hphantom{-\vor \sum_{f=\{f_\uparrow\}} N_C^f \Big[} + 4 
h_{f_\uparrow} m_{f_\uparrow} h_{f_\downarrow} m_{f_\downarrow} 
\sin\b\cos\b\, \dot B_0 \Big] ( m_{H^+}^2, m_{f_\downarrow}^2, 
m_{f_\uparrow}^2) \Big] 
\\ \non
\dot\Pi_{11}^{H^+,\ti\chi} &=& -\vor\, g^2 \sum_{m=1}^4 \sum_{n=1}^2 
\Big[ \Big( \big(F^L_{nm1}\big)^2 + \big(F^R_{nm1}\big)^2 \Big) \Big) 
\Big( \big( m_{\nt_m}^2 + m_{\chp_n}^2 - m_{H^+}^2 \big) \dot B_0 - B_0 
\Big) 
\\
&&\hphantom{-\vor\, g^2 \sum_{m=1}^4 \sum_{n=1}^2 \Big[} + 4 m_{\nt_m} 
m_{\chp_n} F^L_{nm1} F^R_{nm1} \dot B_0 \Big] (m_{H^+}^2, m_{\nt_m}^2, 
m_{\chp_n}^2) \Big] 
\\ 
\dot\Pi_{11}^{H^+,\sf} &=& \vor \sum_{f=\{f_\uparrow\}} N_C^f \sum_{m,n 
= 1}^2 \big(G^{\uparrow\downarrow}_{nm1} \big)^2 \dot B_0 (m_{H^+}^2, 
m_{\sf^{\downarrow}_m}^2, m_{\sf^{\uparrow}_n}^2) 
\\ \non
\dot\Pi_{11}^{H^+,H} &=& \vor \sum_{m,n = 1}^2 \!\Big[ (-1)^n \frac{g 
m_W}{2} (1 \!+\! \d_{1n}) \ti A_{nm}'^{(1)} + \frac{g_Z m_Z}{2} 
R_{2m}(\a\!+\!\b) \ti C_{1n} \Big]^2 
\\ \non
&&\hphantom{\vor \sum_{m,n = 1}^2} \times \dot B_0 (m_{H^+}^2, 
m_{h_m^0}^2, m_{H_n^+}^2) 
\\
&& +\vor \left( \frac{g\,m_W}{2} \right)^2 \dot B_0 (m_{H^+}^2, 
m_{A^0}^2, m_W^2) 
\\ \non
\end{eqnarray}
\begin{eqnarray}\non
\dot\Pi_{11}^{H^+,HV}\!\!\! &=&\!\! -\vor e_0^2 \Big[ \big( 4 m_{H^+}^2 - \l^2 
\big) \dot B_0 + 2 B_0 \Big] (m_{H^+}^2, m_{H^+}^2, \l^2) 
\\ \non
&&\!\! -\vor g_Z^2 \big( \onehfb - s_W^2 \big)^2 \Big[ \big( 4 m_{H^+}^2 - 
m_Z^2 \big) \dot B_0 + 2 B_0 \Big] (m_{H^+}^2, m_{H^+}^2, m_Z^2) 
\\ \non
&&\!\! -\vor \frac{g^2}{4} \sum_{k=1}^2 \big(R_{1k}(\a\!-\!\b)\big)^2 \Big[ 
\big( 2 m_{H^+}^2 + 2 m_{h_k^0}^2 - m_W^2 \big) \dot B_0 + 2 B_0 \Big] 
(m_{H^+}^2, m_{h_k^0}^2, m_W^2) 
\\
&&\!\! -\vor \frac{g^2}{4} \Big[ \big( 2 m_{H^+}^2 + 2 m_{A^0}^2 - m_W^2 
\big) \dot B_0 + 2 B_0 \Big] (m_{H^+}^2, m_{A^0}^2, m_W^2) 
\end{eqnarray}
\subsection{Off-diagonal Wave-function corrections --- \mbox{Mixing} of 
CP-even Higgs bosons} %
According to Section~\ref{renormalization}, the off-diagonal 
wave-function renormalization constants of the external Higgs bosons 
$h_k^0 = \{h^0, H^0\}$ are given by 
\begin{eqnarray}
\d Z_{kl}^{H^0} &=& \displaystyle{\frac{2}{m_{h_k^0}^2 - 
   m_{h_l^0}^2}}\, \Re\, \Pi_{kl}^{H^0}(m_{h_l^0}^2)
   \,,\hspace{2cm} k \neq l
\end{eqnarray}
The single contributions are as follows:
\begin{eqnarray}
\Pi_{12}^{H^0,f}(k^2) &=& -\frac{2}{(4\pi)^2}\, \sum_{f} N_C^f s^f_1 
s^f_2 \Big[ (4 m_f^2 - k^2) B_0(k^2, m_f^2, m_f^2 ) + A_0(m_f^2) )\Big] 
\\ \non
\Pi_{12}^{H^0,\sf}(k^2) &=& \vor\, \sum_{f} \sum_{m,n=1}^2 N_C^f\, 
G^\sf_{nm1} G^\sf_{mn2}\, B_0( k^2, m_{\sf_m}^2, m_{\sf_n}^2 ) 
\\
&& + \vor \sum_{f} \sum_{m=1}^2 N_C^f \big( h_f^2 \,c^\sf_{12} + g^2 
(c^\sb_{12} - c^\st_{12}) \,e^\sf_{mm} \big) A_0(m_{\sf_m}^2) 
\\ \non
\Pi_{12}^{H^0,\nt}(k^2) &=& -\vor\, g^2 \sum_{m,n=1}^4 F^0_{nm1} 
F^0_{mn2} \Big[ \big( (m_{\nt_m}+m_{\nt_n})^2 - k^2 \big) B_0 (k^2, 
m_{\nt_m}^2, m_{\nt_n}^2) 
\\
&&\hphantom{-\vor\, g^2 \sum_{m,n=1}^4 F^0_{nm1} F^0_{mn2} \Big[} + 
A_0(m_{\nt_m}^2) + A_0(m_{\nt_n}^2) \Big] 
\\ \non
\Pi_{12}^{H^0,\chp}(k^2) &=& -\vor\, g^2 \sum_{m,n=1}^2 \Big[ 2 
m_{\chp_m} m_{\chp_n} \Big( F^+_{mn1} F^+_{nm2} + F^+_{nm1} F^+_{mn2} 
\Big) B_0 (k^2, m_{\chp_m}^2, m_{\chp_n}^2) 
\\ \non
&&\hphantom{-\vor\, g^2 \sum_{m,n=1}^2 \Big[} + \Big( F^+_{mn1} 
F^+_{mn2} + F^+_{nm1} F^+_{nm2} \Big) \Big( A_0(m_{\chp_m}^2) + 
A_0(m_{\chp_n}^2) 
\\[2mm]
&&\hphantom{-\vor\, g^2 \sum_{m,n=1}^2 \Big[} + \big( m_{\chp_m}^2 + 
m_{\chp_n}^2 - k^2 \big) B_0(k^2, m_{\chp_m}^2, m_{\chp_n}^2) \Big) 
\Big] 
\end{eqnarray}
\begin{figure}[h!]
\begin{picture}(165,105)(0,0)
     \put(20,0){\mbox{\resizebox{12.5cm}{!}
     {\includegraphics{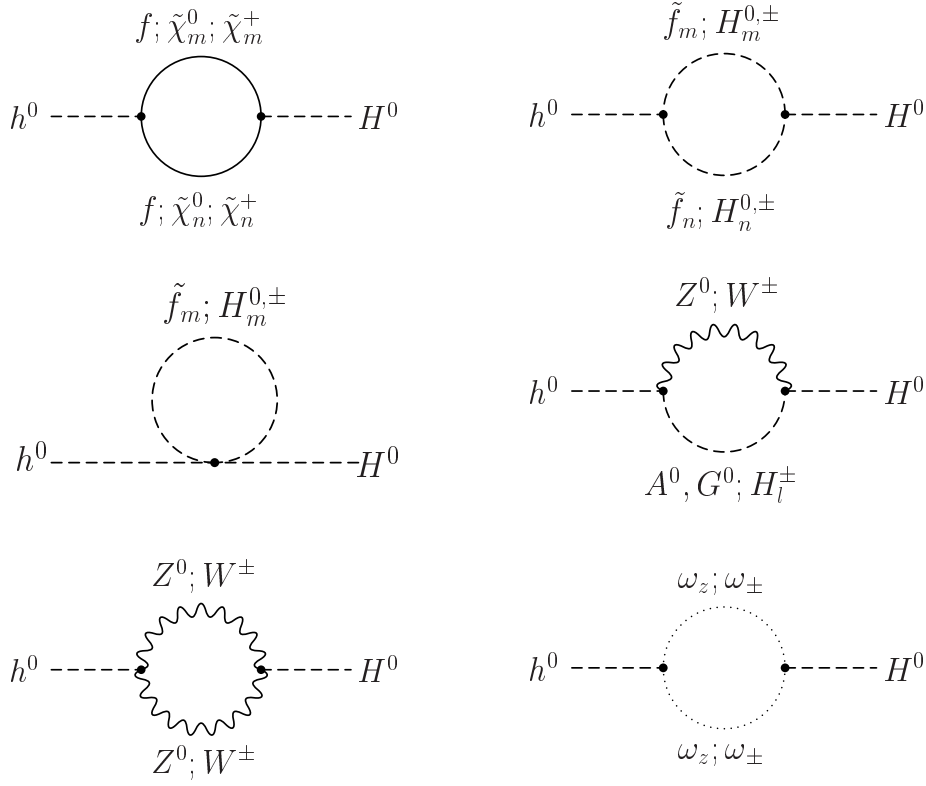}}}}
\end{picture}
\caption{Diagrams for off-diagonal mixing of the CP-even Higgs bosons 
$h^0$ and $H^0$\label{h0H0SE}} 
\end{figure}
%
\begin{eqnarray}\non
\Pi_{12}^{H^0,H}(k^2) &=& \vor \frac{1}{2} \left(\frac{g_{Z} \, 
m_{Z}}{4}\right)^2 \bigg[ \sum_{m,n = 1}^2 (2\!+\!\d_{1m} \d_{mn})! 
\big( \cos 2\a \,\ti A_{mn}^{(1)} - 2 \sin 2\a\, \ti B_{mn}^{(1)} \big) 
\\ \non
&&\hphantom{\vor \frac{1}{2} \left(\frac{g_{Z} \, m_{Z}}{4}\right)^2 
\bigg[} \times (2\!+\!\d_{2m} \d_{mn})! \big( \cos 2\a \,\ti 
A_{mn}^{(2)} - 2 \sin 2\a\, \ti B_{mn}^{(2)} \big) B_0 
\\ \non
&&\hphantom{\vor \left(\frac{g_{Z} \, m_{Z}}{4}\right)^2 
\bigg[} - 2 \sum_{m,n = 3}^4 \sin(2\a + 2\b) \, \big(\ti 
C_{m-2,n-2}\big)^2 B_0 \bigg] (k^2, m_{H_m^0}^2, m_{H_n^0}^2) 
\\ \non
&& \hspace{-0.5cm}+\vor \sum_{m,n = 1}^2 \Big( (-1)^{mn}\, \frac{g \, m_{W}}{2} 
(1-\d_{m2}\d_{n2}) (1+\d_{mn}) \ti A_{mn}'^{(1)} - \frac{g_{Z} \, 
m_{Z}}{2} s_{\a+\b}\, \ti C_{mn} \Big) 
\\ \non
&&\hspace{-0.5cm}\hphantom{\vor x} \times \Big( (-1)^{mn}\, \frac{g \, m_{W}}{2} 
(1-\d_{m2}\d_{n2}) (1+\d_{mn}) \ti A_{mn}'^{(2)} - \frac{g_{Z} \, 
m_{Z}}{2} s_{\a+\b-\pi/2}\, \ti C_{mn} \Big) 
\\[2mm] \non
&&\hspace{-0.5cm}\hphantom{\vor x} \times B_0 (k^2, m_{H_m^+}^2, m_{H_n^+}^2) 
\\ \non
&&\hspace{-0.5cm} +\vor \frac{g_Z^2}{8} \sin 2\a \Big( 3 \big( A_0(m_{h^0}^2) - 
A_0(m_{H^0}^2) \big) \cos 2\a 
\\ \non
&&\hspace{-0.5cm}\hphantom{+\vor \frac{g_Z^2}{8} \sin 2\a \Big(} +\big( A_0(m_{A^0}^2) 
- A_0(m_Z^2) \big) \cos 2\b \Big) 
\\ \non
&&\hspace{-0.5cm} -\vor \Big(\frac{g^2}{2}\sin(\a\!-\!\b) \cos(\a\!-\!\b) - 
\frac{g_Z^2}{2}\sin 2\a \, \cos 2\b \Big) \big( A_0(m_{H^+}^2) - 
A_0(m_W^2) \big) 
\\
\\ \non
\Pi_{12}^{H^0,V}(k^2) &=& -\vor\, \frac{g^2}{2} \sum_{l=1}^2 \Big[ 
\big( 2 k^2 \!+\! 2 m_{H_l^+}^2 \!-\! m_W^2 \big) B_0 (k^2, 
m_{H_l^+}^2, m_W^2) + 2 A_0(m_W^2) 
\\ \non
&&\hspace{-0.5cm}\hphantom{-\vor\, \frac{g^2}{2} \sum_{l=1}^2 \Big[} - 
A_0(m_{H_l^+}^2) + \frac{1}{2 c_W^2}\Big( \big( 2 k^2 \!+\! 2 
m_{H_{l+2}^0}^2 \!-\! m_Z^2 \big) B_0 (k^2, m_{H_{l+2}^0}^2, m_Z^2) 
\\
&&\hspace{-0.5cm}\hphantom{-\vor\, \frac{g^2}{2} \sum_{l=1}^2 \Big[}  + 2 A_0(m_Z^2) - 
A_0(m_{H_{l+2}^0}^2)\Big) \Big] R_{l1}(\a\!-\!\b) R_{l2}(\a\!-\!\b) 
\\ \non
\Pi_{12}^{H^0,VV}(k^2) &=& -\vor \sin(2\a - 2\b) \Big( 2 g^2 m_W^2 B_0 
(k^2, m_W^2, m_W^2) + g_Z^2 m_Z^2 B_0 (k^2, m_Z^2, m_Z^2) \Big) 
\\
\end{eqnarray}
\begin{eqnarray}\non
\Pi_{12}^{H^0,{\rm ghost}}(k^2) &=& \vor \sin(2\a - 2\b) \Big[ 
\frac{g^2}{2} m_W^2 B_0 (k^2, m_W^2, m_W^2) + \frac{g_Z^2}{4} m_Z^2 B_0 
(k^2, m_Z^2, m_Z^2) \Big]  
\\
\end{eqnarray}
\subsection{$H^+ W^+$-mixing\label{appHWmixing}}
\begin{figure}[th]
\begin{picture}(165,105)(0,0)
     \put(5,0){\mbox{\resizebox{16cm}{!}
     {\includegraphics{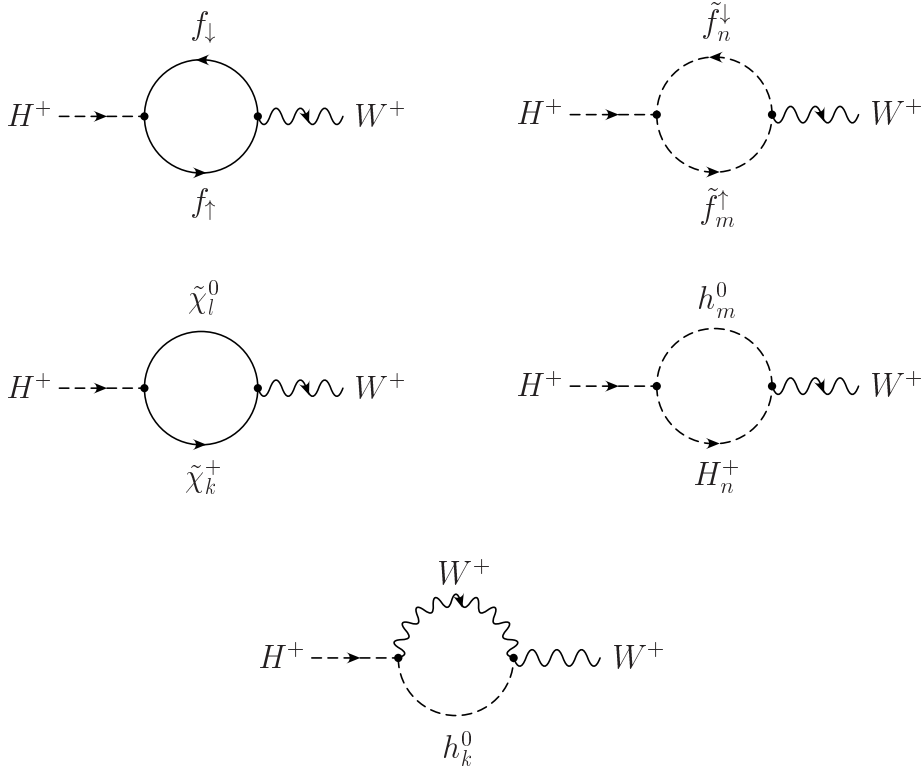}}}}
\end{picture}
\caption{$H^+ W^+$-mixing self-energies\label{HWmixing}} 
\end{figure}

The scalar-vector mixing self-energy, $\Pi_{HW}(k^2)$ (see 
Fig.~\ref{HWmixing}), is defined by the two-point function\\ 
\begin{picture}(140,30)(0,0)
     \put(15,0){\mbox{\resizebox{13cm}{!}
     {\includegraphics{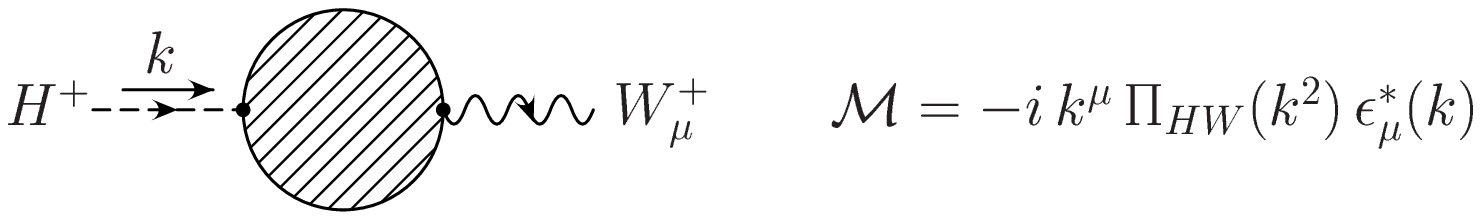}}}}
\end{picture}\\

\begin{eqnarray}\non
\Pi_{HW}^f &=& \vor \, \sqrt2 \, g \sum_{f=\{f_\uparrow\}} N_C^f\, 
\big( m_{f_\downarrow} y^{f_\downarrow}_1 B_1(m_{H^+}^2, 
m_{f_\uparrow}^2, m_{f_\downarrow}^2) - m_{f_\uparrow} y^{f_\uparrow}_1 
B_1(m_{H^+}^2, m_{f_\downarrow}^2, m_{f_\uparrow}^2) \big) 
\\
\\ \non
\Pi_{HW}^{\ti\chi} &=& \vor\, 2\, g^2 \sum_{k=1}^2 \sum_{l=1}^4 \Big[ 
m_{\chp_k} \big( F^L_{kl1} O^{L}_{lk} + F^R_{kl1} O^{R}_{lk} \big) 
\big( B_0 + B_1 \big) 
\\
&&\hphantom{\vor\, 2\, g^2 \sum_{k=1}^2 \sum_{l=1}^4 \Big[} + m_{\nt_l} 
\big( F^L_{kl1} O^{R}_{lk} + F^R_{kl1} O^{L}_{lk} \big) B_1 \Big] 
(m_{H^+}^2, m_{\chp_k}^2, m_{\nt_l}^2) 
\\[2mm]  
\Pi_{HW}^\sf &=& -\vor\, \frac{g}{\sqrt2} \sum_{f=\{f_\uparrow\}} 
N_C^f\, G^{\uparrow\downarrow}_{mn1} R^{\sf_\uparrow}_{m1} 
R^{\sf_\downarrow}_{n1} \big( B_0 + 2 B_1 \big) (m_{H^+}^2, 
m_{\sf_{\uparrow m}}^2, m_{\sf_{\downarrow n}}^2) 
\\[2mm] \non
\Pi_{HW}^H &=& -\vor\, \frac{g}{4} \sum_{m,n = 1}^2 \Big[ (-1)^n g m_W 
(1 \!+\! \d_{1n}) \ti A_{nm}'^{(1)} + g_Z m_Z R_{2m}(\a\!+\!\b) \ti 
C_{1n} \Big] R_{mn}(\b\!-\!\a) 
\\ 
&&\hphantom{-\vor\, \frac{g}{4} \sum_{m,n = 1}^2} \times \big( B_0 + 2 
B_1 \big) (m_{H^+}^2, m_{H_n^+}^2, m_{h_m^0}^2) 
\\[2mm]
\Pi_{HW}^W &=& -\vor\, \frac{g^2\, m_W}{2} \sum_{k=1}^2 
R_{1k}(\a\!-\!\b) R_{2k}(\a\!-\!\b) \big( B_0 - B_1 \big) (m_{H^+}^2, 
m_{h_k^0}^2, m_W^2) 
\end{eqnarray}
%

\section{Vertex corrections}\label{appVertex}
\subsection{$h_k^0 \sf_i \ {\bar{\!\!\sf}}_{\!\!j}$ vertex\label{hk0vertex}}
\begin{figure}[th]
\begin{picture}(160,215)(0,0)
     \put(0,-2){\mbox{\resizebox{16cm}{!}
     {\includegraphics{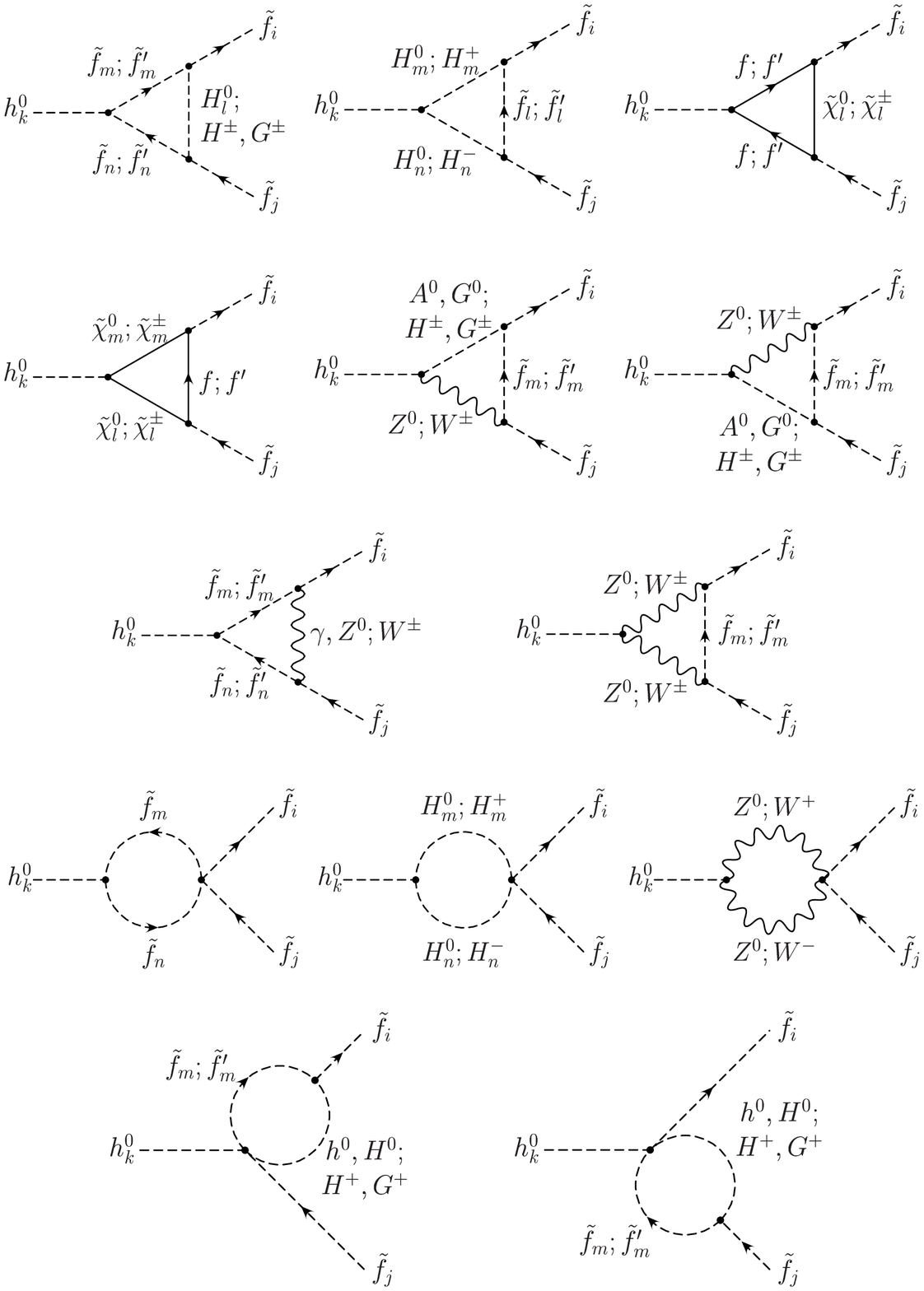}}}}
\end{picture}
\caption{Vertex diagrams relevant to the calculation of the virtual 
electroweak corrections to the decay width $h_k^0 \rightarrow 
\tilde{f}_i \ {\bar{\!\!\tilde{f}}}_{\!j}$ with $h^0_k = \{h^0,H^0\}$. 
\label{vertex-graphshk0}} 
\end{figure}

Here we give the explicit form of the electroweak contributions to the 
vertex corrections which are depicted in Fig.~\ref{vertex-graphshk0}. 
For SUSY-QCD contributions we refer to \cite{SUSY-QCD}. 
\begin{eqnarray}
   \d G_{ijk}^{\sf (v)} &=& \d G_{ijk}^{\sf (v, H\sf\sf)}
      + \d G_{ijk}^{\sf (v, \sf HH)}
      + \d G_{ijk}^{\sf (v, \ti\chi ff)}
      + \d G_{ijk}^{\sf (v, f \ti\chi \ti\chi)}
      + \d G_{ijk}^{\sf (v, VSS)}
   \non \\[3mm]
   && +~\d G_{ijk}^{\sf (v, VVS)}
      + \d G_{ijk}^{\sf (v, SS)}
      + \d G_{ijk}^{\sf (v, VV)}
      + \d G_{ijk}^{\sf (v, hH\rm mix)}
      + \d G_{ijk}^{\sf (v, \sf\rm mix)}
\end{eqnarray}
\noindent The single contributions correspond to the diagrams with 
three scalar particles $( \d G_{ijk}^{\sf (v, H\sf\sf)}$ and $\d 
G_{ijk}^{\sf (v, HH\sf)} )$, three fermions $( \d G_{ijk}^{\sf (v, 
\ti\chi ff)}$ and $\d G_{ijk}^{\sf (v, f \ti\chi \ti\chi)} )$, three 
particles with one or two vector bosons $( \d G_{ijk}^{\sf (v, VSS)}$ 
and $\d G_{ijk}^{\sf (v, VVS)} )$ and two scalar or two vector 
particles $( \d G_{ijk}^{\sf (v, SS)}$ and $\d G_{ijk}^{\sf (v, VV)} )$ 
in the loop. The vertex corrections due to the mixing of the outer 
particles, i.~e. Higgs and sfermion mixing terms $\d G_{ijk}^{\sf (v, 
hH\rm mix)}$ and $\d G_{ijk}^{\sf (v, \sf\rm mix)}$ will be combined 
with the counter terms of the Higgs and sfermion mixing angles, $\d\a$ 
and $\d\theta_\sf$, see eq.~(\ref{symmWF}). 
\noindent The vertex corrections from the exchange of one Higgs boson 
and two sfermions are 
\begin{eqnarray}\non
\d G_{ijk}^{\sf (v, H\sf\sf)} &=& - \frac{1}{(4\pi)^2} \sum_{m,n =
1}^2 \sum_{l=1}^4 G_{mnk}^\sf G_{iml}^\sf G_{njl}^\sf\, C_0\Big(
m_{\sf_i}^2, m_{h_k^0}^2, m_{\sf_j}^2, m_{H_l^0}^2, m_{\sf_m}^2,
m_{\sf_n}^2 \Big)
\\[2mm] \non
&& - \frac{1}{(4\pi)^2} \sum_{m,n = 1}^2 \sum_{l=1}^2
G_{mnk}^{\sf'} G_{iml}^{\sf\sf'} G_{jnl}^{\sf\sf'}\, C_0\Big(
m_{\sf_i}^2, m_{h_k^0}^2, m_{\sf_j}^2, m_{H_l^+}^2, m_{\sf'_m}^2,
m_{\sf'_n}^2 \Big)
\\[-3mm]
\\[-7mm] \non
\end{eqnarray}
with the standard two-point function $C_0$ \cite{PaVe} for which we 
follow the conventions of \cite{Denner}. The graph with 2 Higgs 
particles and one sfermion in the loop leads to 
\begin{eqnarray}\non
\d G_{ijk}^{\sf (v, \sf HH)} &=& -\vor \frac{g_{Z} \, m_{Z}}{4} 
\sum_{m,n = 1}^2 \sum_{l = 1}^2 (2\!+\!\d_{km} \d_{mn})! \left( \cos 
2\a \,\ti A_{mn}^{(k)} - 2 \sin 2\a\, \ti B_{mn}^{(k)} \right) \times 
\\[2mm]\non
&& \hphantom{-\vor \frac{g_{Z} \, m_{Z}}{4} \sum_{m,n = 1}^2 \sum_{l = 
1}^2} G^\sf_{ilm} \, G^\sf_{ljn} \, C_0\Big( m_{\sf_i}^2, m_{h_k^0}^2, 
m_{\sf_j}^2, m_{\sf_l}^2, m_{h_m^0}^2, m_{h_n^0}^2 \Big) 
\\ \non
&& +\vor \frac{g_{Z} \, m_{Z}}{2} \sum_{m,n = 3}^4 \sum_{l = 1}^2 
\sin[\a+\b-\frac{\pi}{2}(k-1)] \, \ti C_{m-2,n-2} \times 
\\ \non
&& \hphantom{+\vor \frac{g_{Z} \, m_{Z}}{2} \sum_{m,n = 3}^4 \sum_{l = 
1}^2} G^\sf_{ilm} \, G^\sf_{ljn} \, C_0\Big( m_{\sf_i}^2, m_{h_k^0}^2, 
m_{\sf_j}^2, m_{\sf_l}^2, m_{H_m^0}^2, m_{H_n^0}^2 \Big) 
\\ \non
&& -\vor \sum_{m,n = 1}^2 \sum_{l = 1}^2 \bigg[ (-1)^{mn}\, \frac{g \, 
m_{W}}{2} (1\!-\!\d_{m2}\d_{n2}) (1\!+\!\d_{mn}) \ti A_{mn}'^{(k)} 
\\ \non
&& \hphantom{-\vor \sum_{m,n = 1}^2 \sum_{l = 1}^2 \bigg[} - 
\frac{g_{Z} \, m_{Z}}{2} \sin[\a+\b-\frac{\pi}{2}(k-1)] \, \ti C_{mn} 
\bigg] \times 
\\
&& \hphantom{-\vor \sum_{m,n = 1}^2 \sum_{l = 1}^2 \bigg[} 
G^{\sf\sf'}_{ilm}\, G^{\sf\sf'}_{jln}\, C_0\Big( m_{\sf_i}^2, 
m_{h_k^0}^2, m_{\sf_j}^2, m_{\sf'_l}^2, m_{H_m^+}^2, m_{H_n^+}^2 \Big) 
\,. 
\end{eqnarray}
For the gaugino exchange contributions we get
\begin{eqnarray}\non
\d G_{ijk}^{\sf (v, \ti\chi ff)} &=& \frac{1}{(4\pi)^2} \sum_{l =
1}^4 F\Big( m_{\sf_i}^2, m_{h_k^0}^2, m_{\sf_j}^2, m_{\nt_l}, m_f,
m_f; s^f_k, s^f_k, b^\sf_{il}, a^\sf_{il}, a^\sf_{jl}, b^\sf_{jl}
\Big)
\\[1mm]\non
&& \hspace{-3mm}+\frac{1}{(4\pi)^2} \sum_{l = 1}^2 F\Big(
m_{\sf_i}^2, m_{h_k^0}^2, m_{\sf_j}^2, m_{\chp_l}, m_{f'}, m_{f'};
s^{f'}_k, -s^{f'}_k, k^\sf_{il}, l^\sf_{il}, l^\sf_{jl},
k^\sf_{jl} \Big)\,,
\\[3mm]\non
\d G_{ijk}^{\sf (v, f \ti\chi \ti\chi)} &=& \frac{1}{(4\pi)^2}
\sum_{l,m = 1}^4 \!\! F\Big( m_{\sf_i}^2, m_{h_k^0}^2,
m_{\sf_j}^2, m_f, m_{\nt_m}, m_{\nt_l}; -g F^0_{lmk}, -g
F^0_{lmk}, b^\sf_{im}, a^\sf_{im}, 
\\[1mm]\non
&& a^\sf_{jl}, b^\sf_{jl} \Big) + 
\frac{1}{(4\pi)^2} \sum_{l,m = 1}^2 \!\! F\Big(
m_{\sf_i}^2, m_{h_k^0}^2, m_{\sf_j}^2, m_{f'}, m_{\chp_m},
m_{\chp_l}; -g \widetilde F^+_{mlk}, -g \widetilde F^+_{lmk},
\\[1mm]
&& k^\sf_{im}, l^\sf_{im}, l^\sf_{jl}, k^\sf_{jl} \Big)\,, 
\end{eqnarray}
where $F(\ldots)$ shortly stands for
\begin{eqnarray}\non
F\Big( m_1^2, m_0^2, m_2^2, M_0, M_1, M_2; g_0^R, g_0^L, g_1^R,
g_1^L, g_2^R, g_2^L \Big) &=& (h_1M_1 \!+\! h_2M_2)B_0(m_0^2,
M_1^2, M_2^2)
\\[2mm]\non
&&\hspace{-9cm} +~(h_0 M_0 \!+\! h_1 M_1)B_0(m_1^2, M_0^2, M_1^2)
+(h_0M_0 \!+\! h_2M_2)B_0(m_2^2, M_0^2, M_2^2)
\\[2mm]\non
&&\hspace{-9cm} +\, \Big[2\!\left(g_0^Rg_1^Rg_2^R \!+\!
g_0^Lg_1^Lg_2^L \right)\!M_0M_1M_2 + h_0M_0\!\left( M_1^2\!
+\!M_2^2\!- \!m_0^2 \right) +
h_1M_1\!\left(M_0^2\!+\!M_2^2\!-\!m_2^2\right)
\\[2mm]
&&\hspace{-9cm}
+~h_2M_2\!\left(M_0^2\!+\!M_1^2\!-\!m_1^2\right)\Big] C_0(m_1^2,
m_0^2, m_2^2, M_0^2, M_1^2, M_2^2)
\end{eqnarray}
with the abbreviations $h_0 = (g_0^Lg_1^Rg_2^R + g_0^Rg_1^Lg_2^L ), h_1 
= (g_0^Lg_1^Lg_2^R + g_0^Rg_1^Rg_2^L )$ and $h_2 = (g_0^Rg_1^Lg_2^R + 
g_0^Lg_1^Rg_2^L )$. For up-type sfermions $\widetilde F^+_{lmk} = 
F^+_{lmk}$ and for down-type sfermions chargino indices are 
interchanged, $\widetilde F^+_{lmk} = F^+_{mlk}$\,. 
\\
We split the irreducible vertex graphs with one vector particle in the 
loop into the single contributions of the photon, the $Z$-boson and the 
$W$-boson, 
\begin{eqnarray}
   \d G_{ijk}^{\sf (v, VSS)} &=& \d G_{ijk}^{\sf (v, \g SS)}
   + \d G_{ijk}^{\sf (v, ZSS)} + \d G_{ijk}^{\sf (v, WSS)}\,.
\end{eqnarray}
In order to regularize the infrared divergences we introduce a
photon mass $\l$. Thus we have
\begin{eqnarray}
\d G_{ijk}^{\sf (v, \g SS)} &=& \frac{1}{(4\pi)^2}\, (e_0 e_f)^2
G^\sf_{ijk}\ V\Big( m_{\sf_i}^2, m_{h_k^0}^2, m_{\sf_j}^2, \l^2,
m_{\sf_i}^2, m_{\sf_j}^2 \Big) \,,
\\ \non
\d G_{ijk}^{\sf (v, ZSS)} &=& \frac{1}{(4\pi)^2}\, g_{Z}^2 
\sum_{m,n = 1}^2 G^\sf_{mnk}\, z^\sf_{im}\, z^\sf_{nj}\ V\Big( 
m_{\sf_i}^2, m_{h_k^0}^2, m_{\sf_j}^2, m_{Z}^2, m_{\sf_m}^2, 
m_{\sf_n}^2 \Big) 
\\ \non
&& \hspace{-3mm}+\frac{i}{(4\pi)^2}\, \frac{g_{Z}^2}{2} \sum_{l =
3}^4 \sum_{m = 1}^2 G^\sf_{mjl}\, z^\sf_{im}\,
R_{l-2,k}(\a\!-\!\b)\ V\Big( m_{h_k^0}^2, m_{\sf_j}^2,
m_{\sf_i}^2, m_{Z}^2, m_{H_l^0}^2, m_{\sf_m}^2 \Big)
\\ \non
&& \hspace{-3mm}-\frac{i}{(4\pi)^2}\, \frac{g_{Z}^2}{2} \sum_{l =
3}^4 \sum_{m = 1}^2 G^\sf_{iml}\, z^\sf_{mj}\,
R_{l-2,k}(\a\!-\!\b)\ V\Big( m_{\sf_j}^2, m_{\sf_i}^2,
m_{h_k^0}^2, m_{Z}^2, m_{\sf_m}^2, m_{H_l^0}^2 \Big) \,,
\\ 
\\[3mm] \non
\d G_{ijk}^{\sf (v, WSS)} &=& \frac{1}{(4\pi)^2}\, \frac{g^2}{2} 
\sum_{m,n = 1}^2 G^{\sf'}_{mnk}\, R^\sf_{i1} R^\sf_{j1} 
R^{\sf'}_{m1} R^{\sf'}_{n1} \ V\Big( m_{\sf_i}^2, m_{h_k^0}^2, 
m_{\sf_j}^2, m_{W}^2, m_{\sf'_m}^2, m_{\sf'_n}^2 \Big) 
\\ \non
&& \hspace{-6mm}-\frac{I_f^{3L}}{(4\pi)^2}\, \frac{g^2}{\sqrt2} 
\sum_{m,l = 1}^2 G^{\sf\sf'}_{jml}\, R^\sf_{i1} R^{\sf'}_{m1}\, 
R_{l,k}(\a\!-\!\b) \ V\Big( m_{h_k^0}^2, m_{\sf_j}^2, m_{\sf_i}^2, 
m_{W}^2, m_{H_l^+}^2, m_{\sf'_m}^2 \Big) 
\\ \non
&& \hspace{-6mm}-\frac{I_f^{3L}}{(4\pi)^2}\, \frac{g^2}{\sqrt2} 
\sum_{m,l = 1}^2 G^{\sf\sf'}_{iml}\, R^{\sf'}_{m1} R^\sf_{j1}\, 
R_{l,k}(\a\!-\!\b) \ V\Big( m_{\sf_j}^2, m_{\sf_i}^2, m_{h_k^0}^2, 
m_{W}^2, m_{\sf'_m}^2, m_{H_l^+}^2 \Big) \,, 
\\
\end{eqnarray}
where we have used the vector vertex function
\begin{eqnarray}\non
V\Big( m_1^2, m_0^2, m_2^2, M_0^2, M_1^2, M_2^2 \Big) \!\!&=&\!\!
- B_0\Big( m_0^2, M_1^2, M_2^2\Big) \!+\! B_0\Big( m_1^2, M_0^2,
M_1^2\Big) \!+\!B_0\Big( m_2^2, M_0^2, M_2^2\Big)
\\[1mm]
&&\hspace{-35mm} +~\left( -2m_0^2 \!+\! m_1^2 \!+\! m_2^2 \!-\!
M_0^2 \!+\! M_1^2 \!+\! M_2^2 \right) C_0\Big( m_1^2, m_0^2, m_2^2, 
M_0^2, M_1^2, M_2^2 \Big)\,. 
\end{eqnarray}
The vertex corrections coming from loops with two vector bosons 
and one sfermion are given by 
\begin{eqnarray} \d G_{ijk}^{\sf (v, VVS)} \!&=&\! 
\d G_{ijk}^{\sf (v, ZZ\sf)} + \d G_{ijk}^{\sf (v, WW\sf')} 
\end{eqnarray}
with
\begin{eqnarray}\non
\d G_{ijk}^{\sf (v, ZZ\sf)} &=& -\frac{1}{(4\pi)^2}\, 
\frac{g_{Z}^3 m_{Z}}{2}\, R_{2k}(\a\!-\!\b) \sum_{m = 1}^2 \bigg[ 
4 B_0\Big( m_{h_k^0}^2, m_{Z}^2, m_{Z}^2\Big) \!-\! B_0\Big( 
m_{\sf_i}^2, m_{\sf_m}^2, m_{Z}^2 \Big) 
\\[1mm] \non
&& \hspace{18mm} -~B_0\Big( m_{\sf_j}^2, m_{\sf_m}^2, m_{Z}^2 
\Big) -\left( m_{h_k^0}^2 \!-\! 2 m_{\sf_i}^2 \!-\! 2 m_{\sf_j}^2 
\!-\! 4 m_{\sf_m}^2 \!+\! 2 m_{Z}^2 \right) \times 
\\[1mm] 
&& \hspace{21mm} C_0\Big( m_{\sf_i}^2, m_{h_k^0}^2, m_{\sf_j}^2,
m_{\sf_m}^2, m_{Z}^2, m_{Z}^2 \Big) \bigg] z^\sf_{im}\, z^\sf_{mj}
\end{eqnarray}
and 
\begin{eqnarray}\non
\d G_{ijk}^{\sf (v, WW\sf')} &=& -\frac{1}{(4\pi)^2}\, \frac{g^3
m_{W}}{4}\, R_{2k}(\a\!-\!\b) \sum_{m = 1}^2 \bigg[ 4 B_0\Big(
m_{h_k^0}^2, m_{W}^2, m_{W}^2\Big) \!-\! B_0\Big( m_{\sf_i}^2,
m_{\sf'_m}^2, m_{W}^2 \Big)
\\[1mm] \non
&& \hspace{18mm} -~B_0\Big( m_{\sf_j}^2, m_{\sf'_m}^2, m_{W}^2 
\Big) -\left( m_{h_k^0}^2 \!-\! 2 m_{\sf_i}^2 \!-\! 2 m_{\sf_j}^2 
\!-\! 4 m_{\sf'_m}^2 \!+\! 2 m_{W}^2 \right) \times 
\\[1mm]
&& \hspace{21mm} C_0\Big( m_{\sf_i}^2, m_{h_k^0}^2, m_{\sf_j}^2,
m_{\sf'_m}^2, m_{W}^2, m_{W}^2 \Big) \bigg] R^\sf_{i1} R^\sf_{j1} \Big( 
R^{\sf'}_{m1} \Big)^2\,. 
\end{eqnarray}
We split the irreducible vertex graphs with two scalar particles into 
the contributions from two Higgs bosons, two sfermions and the 
corrections stemming from Higgs-sfermion loops. For better reading we 
introduce the abbreviations
\begin{equation}\label{4sfconvention}
\begin{array}{l@{\,, \qquad\qquad}l}
   R^{\sf_D}_{ijkl} ~\equiv~ R^\sf_{i1} R^\sf_{j1} R^\sf_{k2} R^\sf_{l2} &
   R_{ijkl}^{\sf\sf'_D} ~\equiv~ R^{\sf}_{i1} R^{\sf}_{j1} R^{\sf'}_{k2} R^{\sf'}_{l2}\,,
   \\[1mm]
   R^{\sf_L}_{ijkl} ~\equiv~ R^\sf_{i1} R^\sf_{j1} R^\sf_{k1} R^\sf_{l1} &
   R_{ijkl}^{\sf\sf'_L} ~\equiv~ R^{\sf}_{i1} R^{\sf}_{j1} R^{\sf'}_{k1} R^{\sf'}_{l1}\,,
   \\[1mm]
   R^{\sf_R}_{ijkl} ~\equiv~ R^\sf_{i2} R^\sf_{j2} R^\sf_{k2} R^\sf_{l2} &
   R_{ijkl}^{\sf\sf'_R} ~\equiv~ R^{\sf}_{i2} R^{\sf}_{j2} R^{\sf'}_{k2} R^{\sf'}_{l2}\,,
   \\[1mm]
   R^{\sf\ \!\hat{\!\!\sf}_{\!F}}_{ijkl} ~\equiv~ R^\sf_{i1} R^\sf_{j2} R^{\ \!\hat{\!\!\sf}}_{k1} 
   R^{\ \!\hat{\!\!\sf}}_{l2} & 
   R^{\ \!\hat{\!\!\sf}\sf_{F}}_{ijkl} ~\equiv~ R^{\ \!\hat{\!\!\sf}}_{i1} 
   R^{\ \!\hat{\!\!\sf}}_{j2} R^\sf_{k1} R^\sf_{l2}\,.
\end{array}
\end{equation}
All other combinations occurring are obvious.
\begin{eqnarray}\non
\d G_{ijk}^{\sf (v, SS)} \!&=&\! \d G_{ijk}^{\sf (v, HH)} + \d 
G_{ijk}^{\sf (v, \sf\sf)} + \d G_{ijk}^{\sf (v, \sf'\sf')} + \d 
G_{ijk}^{\sf (v, \ \hat{\!\!\sf}\ \hat{\!\!\sf})} + \d G_{ijk}^{\sf (v, 
\ \hat{\!\!\sf}'\, \hat{\!\!\sf}')} 
\\[1mm]
&& +\d G_{ijk}^{\sf (v, \ti F \ti F)} + \d G_{ijk}^{\sf (v, \ti F' \ti 
F')} + \d G_{ijk}^{\sf (v,\, \hat{\!\ti F}\, \hat{\!\ti F})} + \d 
G_{ijk}^{\sf (v,\, \hat{\!\ti F}'\, \hat{\!\ti F}')} + \d G_{ijk}^{\sf 
(v, H\sf)} 
\end{eqnarray} 
with 
\begin{eqnarray}\non
\d G_{ijk}^{\sf (v, HH)} \!&=&\! -\vor \frac{g_{Z} \, m_{Z}}{8} 
\sum_{m,n = 1}^2 (2\!+\!\d_{km} \d_{mn})! \left( \cos 2\a \,\ti 
A_{mn}^{(k)} - 2 \sin 2\a\, \ti B_{mn}^{(k)} \right) \times 
\\ \non
&& \hphantom{-\vor \frac{g_{Z} \, m_{Z}}{8} \sum_{m,n = 1}^2} \Big( 
h_f^2 c^\sf_{mn} \d_{ij} + g^2 (c^\sb_{mn}-c^\st_{mn}) e^\sf_{ij} \Big) 
B_0 \Big( m_{h_k^0}^2, m_{h_m^0}^2, m_{h_n^0}^2 \Big) 
\\ \non
&& +\vor \frac{g_{Z} \, m_{Z}}{4} \sum_{m,n = 3}^4
\sin[\a+\b-\frac{\pi}{2}(k-1)] \, \ti C_{m-2,n-2} \times 
\\ \non
&& \hphantom{+\vor \frac{g_{Z} \, m_{Z}}{4} \sum_{m,n = 3}^4} \Big( 
h_f^2 c^\sf_{mn} \d_{ij} + g^2 (c^\sb_{mn}-c^\st_{mn}) e^\sf_{ij} \Big) 
B_0 \Big( m_{h_k^0}^2, m_{H_m^0}^2, m_{H_n^0}^2 \Big) 
\\ \non
&& -\vor \sum_{m,n = 1}^2 \Big( h_{f'}^2 
d^{\sf'}_{mn}R^\sf_{i1}R^\sf_{j1} + h_f^2 d^\sf_{mn} 
R^\sf_{i2}R^\sf_{j2} + g^2 (d^\sb_{mn}-d^\st_{mn}) f^\sf_{ij} \Big) 
\times 
\\ \non
&& \hphantom{-\vor \sum_{m,n = 1}^2} \bigg[ (-1)^{mn}\, \frac{g \, 
m_{W}}{2} (1\!-\!\d_{m2}\d_{n2}) (1\!+\!\d_{mn}) \ti A_{mn}'^{(k)} 
\\ \non
&& \hphantom{-\vor \sum_{m,n = 1}^2 \bigg[} - \frac{g_{Z} \, m_{Z}}{2} 
\sin[\a+\b-\frac{\pi}{2}(k-1)] \, \ti C_{mn} \bigg] B_0 \Big( 
m_{h_k^0}^2, m_{H_m^+}^2, m_{H_n^+}^2 \Big) \,, 
\\
\end{eqnarray} 
\begin{eqnarray}\non
\d G_{ijk}^{\sf (v, \sf\sf)} \!&=&\! -\frac{h_f^2}{(4\pi)^2} \sum_{m,n=1}^2 
\!\! G_{nmk}^\sf \bigg[ R^\sf_{ijmn} \!+\! R^\sf_{mnij} \!+\! N_C^f 
\Big( R^\sf_{inmj} \!+\! R^\sf_{mjin} \Big)\! \bigg] B_0 \Big( 
m_{h_k^0}^2, m_{\sf_m}^2, m_{\sf_n}^2 \Big) 
\\[1mm] \non
&& -\frac{g_{Z}^2}{(4\pi)^2} \sum_{m,n=1}^2 \!\! G_{nmk}^\sf \Bigg\{ \bigg[ 
\Big( \frac{1}{4} - (2 I_f^{3L} \!-\! e_f) e_f s_{W}^2 \Big) R^{\sf_L}_{ijmn} + 
e_f^2 s_{W}^2 R^{\sf_R}_{ijmn} \bigg] (N_C^f + 1) 
\\ \non
&&\hspace{35mm} + (I_f^{3L} \!-\! e_f) e_f s_{W}^2 \bigg[ N_C^f \Big( 
R^{\sf}_{ijmn} \!+\! R^{\sf}_{mnij} \Big) \!+\! R^{\sf}_{inmj} \!+\! 
R^{\sf}_{mjin} \bigg] \! \Bigg\} 
\\[1mm]
&&\hspace{35mm} \times \, B_0 \Big( m_{h_k^0}^2, m_{\sf_m}^2, m_{\sf_n}^2 
\Big)\,, 
\end{eqnarray} 
\begin{eqnarray}\non
\d G_{ijk}^{\sf (v, \sf'\sf')} \!&=&\! -\vor \sum_{m,n=1}^2 \!\! 
G_{nmk}^{\sf'} \, \Bigg\{ h_f^2 \, R^{\sf'\!\sf_D}_{mnij} + h_{f'}^2\, 
R^{\sf\sf'_D}_{ijmn} + \frac{g^2}{4} \Bigg[ N_C^f \bigg( \Big( t_{W}^2 
Y^f_L Y^{f'}_L \!-\! 1 \Big) R^{\sf\sf'_L}_{ijmn} 
\\[1mm] \non
&&\hspace{35mm}+ t_{W}^2 Y^f_R Y^{f'}_R \, R^{\sf\sf'_R}_{ijmn} - t_{W}^2 Y^f_L 
Y^{f'}_R R^{\sf\sf'_D}_{ijmn} - t_{W}^2 Y^{f'}_L Y^f_R 
R^{\sf'\!\sf_D}_{mnij} \bigg) 
\\[1mm]
&&\hspace{35mm}  + 2 R^{\sf\sf'_L}_{ijmn} \Bigg] \Bigg\} B_0 \Big( m_{h_k^0}^2, 
m_{\sf'_m}^2, m_{\sf'_n}^2 \Big)\,, 
\end{eqnarray} 
\begin{eqnarray}\label{dGvhatsfhatsf}\non
\d G_{ijk}^{\sf (v, \ \hat{\!\!\sf}\ \hat{\!\!\sf})} \!&=&\! -\frac{N_C^{\hat 
f}}{(4\pi)^2} \sum_{m,n=1}^2 \!\! G_{nmk}^{\ \hat{\!\!\sf}} \, \Bigg\{ 
h_f h_{\hat f} \Big( R^{\sf \ \hat{\!\!\sf}_{F}}_{jimn} + R^{\sf 
\ \hat{\!\!\sf}_{F}}_{ijnm} \Big) + \frac{g^2}{4} \Bigg[ \Big( t_{W}^2 Y^f_L 
Y^{\hat f}_L \!+\! 1 \Big) R^{\sf\ \hat{\!\!\sf}_{L}}_{ijmn} 
\\[1mm] \non
&&\hspace{35mm}+ t_{W}^2 Y^f_R Y^{\hat f}_R \, 
R^{\sf\ \hat{\!\!\sf}_{R}}_{ijmn}  - t_{W}^2 Y^f_L Y^{\hat f}_R R^{\sf\ 
\hat{\!\!\sf}_{D}}_{ijmn}  - t_{W}^2 Y^{\hat f}_L Y^f_R R^{\ 
\hat{\!\!\sf}\sf_D}_{mnij} \Bigg] \Bigg\} 
\\[1mm]
&&\hspace{35mm} \times B_0 \Big( m_{h_k^0}^2, m_{\ \hat{\!\!\sf}_m}^2, 
m_{\ \hat{\!\!\sf}_n}^2 \Big)\,, 
\end{eqnarray} 
and 
\begin{eqnarray}\label{dGvhatsfphatsfp}\non
\d G_{ijk}^{\sf (v, \ \hat{\!\!\sf}' \ \hat{\!\!\sf}')} \!&=&\! -\frac{N_C^{\hat 
f}}{(4\pi)^2} \frac{g^2}{4} \sum_{m,n=1}^2 \!\! G_{nmk}^{\,\ \hat{\!\!\sf}'} \, 
\Bigg\{ \Big( t_{W}^2 Y^f_L Y^{\hat{f}'}_L \!-\! 1 \Big) R^{\sf\ 
\hat{\!\!\sf}'_{L}}_{ijmn} + t_{W}^2 Y^f_R Y^{\hat{f}'}_R \, R^{\sf\ 
\hat{\!\!\sf}'_{R}}_{ijmn} 
\\[1mm] \non
&&\hspace{28mm} - t_{W}^2 Y^f_L Y^{\hat{f}'}_R 
R^{\sf\ \hat{\!\!\sf}'_{D}}_{ijmn} - t_{W}^2 Y^{\hat{f}'}_L Y^f_R R^{\ 
\hat{\!\!\sf}'\sf_D}_{mnij} \Bigg\} B_0 \Big( m_{h_k^0}^2, m_{\ 
\hat{\!\!\sf}'_{m}}^2, m_{\ \hat{\!\!\sf}'_{n}}^2 \Big)\,. 
\\[1mm]
\end{eqnarray}
The contributions due to the exchange of sfermions from the other 
two generations, $\ti F_m$, are given by 
\begin{eqnarray}
\begin{array}{r@{\ }c@{\ }l@{\qquad\quad}c@{\ }r@{\ }l}
\d G_{ijk}^{\sf (v, \ti F\ti F)} &=& \d G_{ijk}^{\sf (v, \ 
\hat{\!\!\sf}\ \hat{\!\!\sf})} (\hat f \rightarrow F)\,, & \d 
G_{ijk}^{\sf (v,\, \hat{\!\ti F}\, \hat{\!\ti F})} &=& \d 
G_{ijk}^{\sf (v, \ \hat{\!\!\sf}\ \hat{\!\!\sf})} (\hat f 
\rightarrow \hat F)\,, 
\\[3mm]
\d G_{ijk}^{\sf (v, \ti F' \ti F')} &=& \d G_{ijk}^{\sf (v, \ 
\hat{\!\!\sf}'\, \hat{\!\!\sf}')} (\hat f' \rightarrow F')\,, & \d 
G_{ijk}^{\sf (v,\, \hat{\!\ti F}'\, \hat{\!\ti F}')} &=& \d 
G_{ijk}^{\sf (v, \ \hat{\!\!\sf}'\, \hat{\!\!\sf}')} (\hat f' 
\rightarrow \hat F')\,, 
\end{array}
\end{eqnarray}
where $\ti F$ denotes values belonging to 
the scalar fermions with the same isospin as 
$\sf$, but from the other two generations (e.g. $\ti F_1 = \{ \ti u_1, \ti c_1 \}$ 
for the stop case, \ldots), and $\ti F'$ sfermions with different isospin etc. \\%
With the abbreviations 
\begin{eqnarray}\non
c_{kl}^{\st,0+} &=& \left( 
\begin{array}{cccc}
    \cos\alpha \cos\beta & \cos\alpha \sin\beta & ~~0~~ & 0
    \\
    \sin\alpha \cos\beta & \sin\alpha \sin\beta & 0 & 0
    \\
    -i \cos^2\beta & -\frac{i}{2}\sin 2\beta & 0 & 0
    \\
    -\frac{i}{2}\sin 2\beta & -i \sin^2\beta & 0 & 0
\end{array}\right)\,,
\\[2mm] \non
c_{kl}^{\sb,0+} &=& \left( 
\begin{array}{cccc}
    0 & ~~0~~ & -\sin\alpha \sin\beta & \sin\alpha \cos\beta
    \\
    0 & 0 & \cos\alpha \sin\beta & -\cos\alpha \cos\beta
    \\
    0 & 0 & -i \sin^2\beta & \frac{i}{2}\sin 2\beta
    \\
    0 & 0 & \frac{i}{2}\sin 2\beta & -i \cos^2\beta
\end{array}\right)\,, 
\\[2mm] \non
c_{kl}^{\st\sb,0+} &=& \left( 
\begin{array}{cccc}
    -\sin\alpha \cos\beta & -\sin\alpha \sin\beta & ~~0~~ & 0
    \\
    \cos\alpha \cos\beta & \cos\alpha \sin\beta & 0 & 0
    \\
    -\frac{i}{2}\sin 2\beta & -i \sin^2\beta & 0 & 0
    \\
    i \cos^2\beta & \frac{i}{2}\sin 2\beta & 0 & 0
\end{array}\right) \,,
\\[2mm] \non
c_{kl}^{\sb\st,0+} &=& \left( 
\begin{array}{cccc}
    0 & ~~0~~ & \cos\alpha \sin\beta & -\cos\alpha \cos\beta
    \\
    0 & 0 & \sin\alpha \sin\beta & -\sin\alpha \cos\beta
    \\
    0 & 0 & -\frac{i}{2}\sin 2\beta & i \cos^2\beta
    \\
    0 & 0 & -i \sin^2\beta & \frac{i}{2}\sin 2\beta
\end{array}\right) 
\end{eqnarray}
the diagrams with one Higgs boson and one sfermion in the loop lead to 
\begin{eqnarray}\non
\d G_{ijk}^{\sf (v, H\sf)} &=& -\frac{1}{(4\pi)^2}\, 
\sum_{l,m=1}^2 G^\sf_{iml} \bigg( h_f^2\, c^\sf_{kl}\, \d_{mj} + 
g^2 \Big( c^\sb_{kl} - c^\st_{kl} \Big) e^\sf_{mj} \bigg) B_0 
\Big( m_{\sf_i}^2, m_{h_l^0}^2, m_{\sf_m}^2 \Big) 
\\[2mm] \non
&&+\frac{1}{(4\pi)^2}\,\frac{1}{\sqrt2} \sum_{l,m=1}^2 \! 
G^{\sf\sf'}_{iml} \bigg( \Big(h_\uparrow^2\, c^{\st,0+}_{kl} + 
h_\downarrow^2\, c^{\sb,0+}_{k,l+2}\Big) R^{\sf}_{j1} R^{\sf'}_{m1} + 
h_\uparrow h_\downarrow \Big( c^{\st\sb,0+}_{kl} + 
c^{\sb\st,0+}_{k,l+2} \Big) 
\\[2mm] \non
&&\hspace{25mm} \times R^{\sf}_{j2} R^{\sf'}_{m2} - \frac{g^2}{2} 
\Big( c^{\st,0+}_{kl} + c^{\sb,0+}_{k,l+2} \Big) R^{\sf}_{j1} 
R^{\sf'}_{m1} \bigg) B_0 \Big( m_{\sf_i}^2, m_{H_l^+}^2, 
m_{\sf'_m}^2 \Big) 
\\[2mm] 
&&\ +\ i \leftrightarrow j \,.
\end{eqnarray}
with $h_\uparrow = \{h_{f_\uparrow}, 0\}$ and $h_\downarrow = 
\{h_{f_\downarrow}, h_{\hat f_\downarrow}\}$ for the decay into 
\{squarks, sleptons\}, respectively. $i \leftrightarrow j$ stands for 
both terms before with $i$ and $j$ interchanged. 
\\%
Finally, for the vertex graphs with two vector bosons we obtain
\begin{eqnarray}\non
\d G_{ijk}^{\sf (v, VV)} &=& \frac{4}{(4\pi)^2}\, g_{Z}^3 m_{Z}\, 
R_{2k}(\a\!-\!\b) \bigg[ \Big(C^f_L\Big)^2 R^\sf_{i1} R^\sf_{j1} + 
\Big(C^f_R\Big)^2 R^\sf_{i2} R^\sf_{j2} \bigg] B_0\Big( 
m_{h_k^0}^2, m_Z^2, m_Z^2 \Big) 
\\[2mm]
&& + \frac{2}{(4\pi)^2}\, g^3 m_W\, R_{2k}(\a\!-\!\b) R^\sf_{i1} 
R^\sf_{j1} B_0\Big( m_{h_k^0}^2, m_W^2, m_W^2 \Big) \,.
\end{eqnarray}

\subsection{$H^+ \sf^\uparrow_i \bar{\sf^\downarrow_j}$ vertex\label{Hpvertex}}
In this chapter we list the various contributions to the vertex 
corrections of the charged Higgs boson decays $H^+ \rightarrow 
\sf^\uparrow_i \bar{\sf^\downarrow_j}$. For simplicity the formulae are 
given for third generation sfermions as the substitution for the first 
and second generation sfermions are obvious. Using the definitions for 
various generic vertex functions and products of couplings from the 
previous chapters we get for the vertex corrections, 
\begin{eqnarray}\non
   \d G_{ij1}^{\st\sb (v)} &=& \d G_{ij1}^{\st\sb (v, H\sf\sf')} 
      + \d G_{ij1}^{\st\sb (v, HH\sf)} 
      + \d G_{ij1}^{\st\sb (v, \ti\chi ff)} 
      + \d G_{ijk}^{\st\sb (v, f \ti\chi \ti\chi)}
      + \d G_{ij1}^{\st\sb (v, \g SS)} 
   \non \\[3mm]
   && +~\d G_{ij1}^{\st\sb (v, ZSS)} 
      + \d G_{ij1}^{\st\sb (v, WSS)} 
      + \d G_{ij1}^{\st\sb (v, HH)} 
      + \d G_{ij1}^{\st\sb (v, \sf\sf')} 
      + \d G_{ij1}^{\st\sb (v, \sF\sF')} 
   \non \\[3mm]
   && +~\d G_{ij1}^{\st\sb (v, H\sf)} 
\end{eqnarray}
The single contributions are given as follows: 
\begin{eqnarray}\non
\d G_{ij1}^{\st\sb (v, H\sf\sf')} &=& - \vor \sum_{m,n = 1}^2 
\sum_{k=1}^4 G_{mn1}^{\st\sb} G_{imk}^{\st} G_{njk}^{\sb}\, C_0\Big( 
m_{{\st}_i}^2, m_{H^+}^2, m_{{\sb}_j}^2, m_{H_k^0}^2, m_{{\st}_m}^2, 
m_{{\sb}_n}^2 \Big) 
\\
\\ \non
\d G_{ij1}^{\st\sb (v, HH\sf)} &=& - \vor \frac{1}{2} \sum_{k,l,m = 
1}^2 \Big[ (-1)^l g\, m_W (1 + \d_{1l}) \ti A_{lk}'^{(1)} + g_Z\, m_Z 
R_{2k}(\a\!+\!\b) \ti C_{1l} \Big] \times 
\\ \non
&& \hspace{3.5cm} G^{\st}_{imk} G^{\st\sb}_{mjl} C_0\Big( 
m_{{\st}_i}^2, m_{H^+}^2, m_{{\sb}_j}^2, m_{{\st}_m}^2, m_{h_k^0}^2, 
m_{H_l^+}^2 \Big) 
\\ \non
&& +\vor \frac{i g m_W}{2} \sum_{m = 1}^2 G^{\st}_{im3} 
G^{\st\sb}_{mj2} C_0\Big( m_{{\st}_i}^2, m_{H^+}^2, m_{{\sb}_j}^2, 
m_{{\st}_m}^2, m_{A^0}^2, m_{W}^2 \Big) 
\\ \non
&& - \vor \frac{1}{2} \sum_{k,l,m = 1}^2 \Big[ (-1)^l g\, m_W (1 + 
\d_{1l}) \ti A_{lk}'^{(1)} + g_Z\, m_Z R_{2k}(\a\!+\!\b) \ti C_{1l} 
\Big] \times 
\\ \non
&& \hspace{3.5cm} G^{\sb}_{mjk} G^{\st\sb}_{iml} C_0\Big( 
m_{{\st}_i}^2, m_{H^+}^2, m_{{\sb}_j}^2, m_{{\sb}_m}^2, m_{H_l^+}^2, 
m_{h_k^0}^2 \Big) 
\\ 
&& +\vor \frac{i g m_W}{2} \sum_{m = 1}^2 G^{\sb}_{mj3} 
G^{\st\sb}_{im2} C_0\Big( m_{{\st}_i}^2, m_{H^+}^2, m_{{\sb}_j}^2, 
m_{{\sb}_m}^2, m_{W}^2, m_{A^0}^2 \Big) 
\end{eqnarray}
\begin{eqnarray} 
\d G_{ij1}^{\st\sb (v, \ti\chi ff)} &=& \vor \sum_{k = 1}^4 F\Big( 
m_{{\st}_i}^2, m_{H^+}^2, m_{{\sb}_j}^2, m_{\nt_k}, m_{t}, m_{b}; 
y^b_1, y^t_1, b^{\st}_{ik}, a^{\st}_{ik}, a^{\sb}_{jk}, b^{\sb}_{jk} 
\Big) 
\\ \non
\d G_{ijk}^{\sf (v, f \ti\chi \ti\chi)} &=& 
\\ \non
&& \hspace{-1cm} -\vor \sum_{k = 1}^4 \sum_{l = 1}^2 F\Big( 
m_{{\st}_i}^2, m_{H^+}^2, m_{{\sb}_j}^2, m_{t}, m_{\nt_k}, m_{\chp_l}; 
-g F^R_{lk1}, -g F^L_{lk1}, b^{\st}_{ik}, a^{\st}_{ik}, l^{\sb}_{jl}, 
k^{\sb}_{jl} \Big) 
\\ \non
&& \hspace{-1cm} -\vor \sum_{k = 1}^4 \sum_{l = 1}^2 F\Big( 
m_{{\st}_i}^2, m_{H^+}^2, m_{{\sb}_j}^2, m_{b}, m_{\chp_l}, m_{\nt_k}; 
-g F^R_{lk1}, -g  F^L_{lk1}, k^{\st}_{il}, l^{\st}_{il}, a^{\sb}_{jk}, 
b^{\sb}_{jk} \Big) 
\\
\end{eqnarray}
\begin{eqnarray}\non
\d G_{ij1}^{\st\sb (v, \g SS)} &=& \vor\, e_0^2 e_t e_b 
G^{\st\sb}_{ij1}\ V\Big( m_{{\st}_i}^2, m_{H^+}^2, m_{{\sb}_j}^2, \l^2, 
m_{{\st}_i}^2, m_{{\sb}_j}^2 \Big) 
\\ \non
&& + \vor\, e_0^2 e_t G^{\st\sb}_{ij1}\ V\Big( m_{H^+}^2, 
m_{{\sb}_j}^2, m_{{\st}_i}^2, \l^2, m_{H^+}^2, m_{{\st}_i}^2 \Big) 
\\
&& - \vor\, e_0^2 e_b G^{\st\sb}_{ij1}\ V\Big( m_{{\sb}_j}^2, 
m_{{\st}_i}^2, m_{H^+}^2, \l^2, m_{{\sb}_j}^2, m_{H^+}^2 \Big) 
\\ \non
\d G_{ij1}^{\st\sb (v, ZSS)} &=& \vor\, g_{Z}^2 \sum_{m,n = 1}^2 
G^{\st\sb}_{mn1}\, z^{\st}_{im}\, z^{\sb}_{nj}\ V\Big( m_{{\st}_i}^2, 
m_{H^+}^2, m_{{\sb}_j}^2, m_Z^2, m_{{\st}_m}^2, m_{{\sb}_n}^2 \Big) 
\\ \non
&& +\vor\, g_{Z}^2 \Big(\onehfb - s_W^2\Big) \sum_{m = 1}^2 
G^{\st\sb}_{mj1}\, z^{\st}_{im}\ V\Big( m_{H^+}^2, m_{{\sb}_j}^2, 
m_{{\st}_i}^2, m_{Z}^2, m_{H^+}^2, m_{{\st}_m}^2 \Big) 
\\ \non
&& -\vor\, g_{Z}^2 \Big(\onehfb - s_W^2\Big) \sum_{m = 1}^2 
G^{\st\sb}_{im1}\, z^{\sb}_{mj}\ V\Big( m_{{\sb}_j}^2, m_{{\st}_i}^2, 
m_{H^+}^2, m_{Z}^2, m_{{\sb}_m}^2, m_{H^+}^2 \Big) 
\\ 
\\ \non
\d G_{ij1}^{\st\sb (v, WSS)} &=& \vor\, \frac{g^2}{2\sqrt2} \sum_{m = 
1}^2 \sum_{k = 1}^3 G^{{\sb}}_{mjk}\, R^{\st}_{i1} R^{{\sb}}_{m1} w_k 
\, V\Big( m_{H^+}^2, m_{{\sb}_j}^2, m_{{\st}_i}^2, m_{W}^2, 
m_{H_k^0}^2, m_{{\sb}_m}^2 \Big) 
\\ \non
&& -\vor\, \frac{g^2}{2\sqrt2} \sum_{m = 1}^2 \sum_{k = 1}^3 
G^{{\st}}_{imk}\, R^{{\st}}_{m1} R^{\sb}_{j1} w_k \, V\Big( 
m_{{\sb}_j}^2, m_{{\st}_i}^2, m_{H^+}^2, m_{W}^2, m_{{\st}_m}^2, 
m_{H_k^0}^2 \Big) \,, 
\\
\end{eqnarray}
with $w_k = \{\cos(\a-\b), \sin(\a-\b), -i\}_k$. 

\begin{eqnarray}\non
\d G_{ij1}^{\st\sb (v, HH)} &=& - \vor \frac{1}{2 \sqrt2} \sum_{k,l = 
1}^2 \Big[ (-1)^l g\, m_W (1 + \d_{1l}) \ti A_{lk}'^{(1)} + g_Z\, m_Z 
R_{2k}(\a\!+\!\b) \ti C_{1l} \Big] \times 
\\ \non
&& \hphantom{- \vor \frac{1}{2 \sqrt2} \sum_{k,l = 1}^2} \Big[ \Big( 
(h_{t}^2 - \frac{g^2}{2}) c_{kl}^{{\st},0+} + (h_{b}^2 - \frac{g^2}{2}) 
(c_{k,l+2}^{{\sb},0+})^\ast \Big) R^{\st}_{i1} R^{\sb}_{j1} 
\\ \non
&& \hphantom{- \vor \frac{1}{2 \sqrt2} \sum_{k,l = 1}^2\,\,} + h_{t} 
h_{b} \Big(c_{kl}^{\st\sb,0+} + (c_{k,l+2}^{{\sb}{\st},0+})^\ast\Big) 
R^{\st}_{i2} R^{\sb}_{j2} \Big] B_0\Big( m_{H^+}^2, m_{h_k^0}^2, 
m_{H_l^+}^2 \Big) 
\\ \non
&& \hspace{-0.5cm}- \vor \frac{g m_W}{2 \sqrt2} \Big[ \onehfb (h_{t}^2 
\!+\! h_{b}^2 \!-\! g^2)\sin 2\b R^{\st}_{i1} R^{\sb}_{j1} + h_{t} 
h_{b} R^{\st}_{i2} R^{\sb}_{j2} \Big] B_0\Big( m_{H^+}^2, m_{A^0}^2, 
m_{W}^2 \Big) 
\\
\end{eqnarray}
\begin{eqnarray}\non
\d G_{ij1}^{\st\sb (v, \sf\sf')} &=& -\vor \sum_{m,n=1}^2 \!\! 
G_{nm1}^{\st\sb} \, \Big\{ N_C^f \Big( h_{t}^2 \, 
R^{{\sb}{\st}_D}_{mjin} + h_{b}^2\, R^{\st\sb_D}_{inmj}\Big) + 
\frac{g^2}{4} \Big[ (2 N_C^f - 1) R^{\st\sb_L}_{inmj} 
\\ \non
&& \hphantom{-\vor \sum_{m,n=1}^2 \!\! G_{nm1}^{\st\sb} \, \Big\{} + 
t_W^2 \Big(Y^t_L Y^{b}_L R^{\st\sb_L}_{inmj} + Y^t_R Y^{b}_R 
R^{\st\sb_R}_{inmj} - Y^t_L Y^{b}_R R^{\st\sb_D}_{inmj} 
\\ \non
&& \hphantom{-\vor \sum_{m,n=1}^2 \!\! G_{nm1}^{\st\sb} \, \Big\{} - 
Y^{b}_L Y^t_R R^{{\sb}{\st}_D}_{mjin}\Big) \Big] \Big\} B_0\Big( 
m_{H^+}^2, m_{{\sb}_m}^2, m_{{\st}_n}^2 \Big) 
\\ \non
&& -\vor \sum_{m=1}^2 G_{1m1}^{\snutau\stau} \Big( h_{b} h_\tau 
\Rsl_{m2} \Rst_{i1} \Rsb_{j2} + \frac{g^2}{2} \Rsl_{m1} \Rst_{i1} 
\Rsb_{j1} \Big) B_0\Big( m_{H^+}^2, m_{{\st}au_m}^2, m_{\snutau}^2 
\Big) 
\\
\end{eqnarray}
\begin{eqnarray}\non
\d G_{ij1}^{\st\sb (v, \sF\sF')} &=& -\frac{N_C^f}{(4\pi)^2} 
\sum_{m,n=1}^2 G_{nm1}^{\su\sd} \Big( \frac{g^2}{2} 
\Rst_{i1}\Rsb_{j1}\Rsu_{n1}\Rsd_{m1} + h_{t} h_u 
\Rst_{i2}\Rsb_{j1}\Rsu_{n2}\Rsd_{m1} 
\\ \non
&& \hphantom{-\frac{N_C^f}{(4\pi)^2} \sum_{m,n=1}^2 G_{nm1}^{\su\sd} 
\Big(} + h_{b} h_d \Rst_{i1}\Rsb_{j2}\Rsu_{n1}\Rsd_{m2} \Big) B_0\Big( 
m_{H^+}^2, m_{\sd_m}^2, m_{\su_n}^2 \Big) 
\\ \non
&& -\vor \sum_{m=1}^2 G_{1m1}^{\sen\se} \Big( \frac{g^2}{2} 
\Rst_{i1}\Rsb_{j1}\Rse_{m1} + h_{b} h_e \Rst_{i1}\Rsb_{j2}\Rse_{m2} 
\Big) B_0\Big( m_{H^+}^2, m_{\se_m}^2, m_{\sen}^2 \Big) 
\\ \non
&& -\frac{N_C^f}{(4\pi)^2} \sum_{m,n=1}^2 G_{nm1}^{\sc\sstrange} \Big( 
\frac{g^2}{2} \Rst_{i1}\Rsb_{j1}\Rsc_{n1}\Rss_{m1} + h_{t} h_c 
\Rst_{i2}\Rsb_{j1}\Rsc_{n2}\Rss_{m1} 
\\ \non
&& \hphantom{-\frac{N_C^f}{(4\pi)^2} \sum_{m,n=1}^2 G_{nm1}^{\su\sd} 
\Big(} + h_{b} h_s \Rst_{i1}\Rsb_{j2}\Rsc_{n1}\Rss_{m2} \Big) B_0\Big( 
m_{H^+}^2, m_{\sstrange_m}^2, m_{\sc_n}^2 \Big) 
\\ \non
&& -\vor \sum_{m=1}^2 G_{1m1}^{\smn\sm} \Big( \frac{g^2}{2} 
\Rst_{i1}\Rsb_{j1}\Rsm_{m1} + h_{b} h_\mu \Rst_{i1}\Rsb_{j2}\Rsm_{m2} 
\Big) B_0\Big( m_{H^+}^2, m_{\sm_m}^2, m_{\smn}^2 \Big) 
\\
\end{eqnarray}
\begin{eqnarray}\non
\d G_{ij1}^{\st\sb (v, H\sf)} &=& -\vor \frac{1}{\sqrt2} \sum_{k=1}^4 
\sum_{m=1}^2 G_{imk}^{{\st}} \Big[ \Rst_{m1}\Rsb_{j1} \Big( h_{t}^2 
c^{{\st},0+}_{k1} + h_{b}^2 (c^{{\sb},0+}_{k3})^\ast \Big) + 
\Rst_{m2}\Rsb_{j2} h_{t} h_{b} \times 
\\ \non
&& \hphantom{\vor} \Big( c^{\st\sb,0+}_{k1} + 
(c^{{\sb}{\st},0+}_{k3})^\ast \Big) - \frac{g^2}{2} \Rst_{m1}\Rsb_{j1} 
\Big( c^{{\st},0+}_{k1} + (c^{{\sb},0+}_{k3})^\ast \Big) \Big] B_0\Big( 
m_{{\st}_i}^2, m_{H_k^0}^2, m_{{\st}_m}^2 \Big) 
\\ \non
&& -\vor \frac{1}{\sqrt2} \sum_{k=1}^4 \sum_{m=1}^2 G_{mjk}^{{\sb}} 
\Big[ \Rst_{i1}\Rsb_{m1} \Big( h_{t}^2 c^{{\st},0+}_{k1} + h_{b}^2 
(c^{{\sb},0+}_{k3})^\ast \Big) + \Rst_{i2}\Rsb_{m2} h_{t} h_{b} \times 
\\ \non
&& \hphantom{\vor} \Big( c^{\st\sb,0+}_{k1} + 
(c^{{\sb}{\st},0+}_{k3})^\ast \Big) - \frac{g^2}{2} \Rst_{i1}\Rsb_{m1} 
\Big( c^{{\st},0+}_{k1} + (c^{{\sb},0+}_{k3})^\ast \Big) \Big] B_0\Big( 
m_{{\sb}_j}^2, m_{H_k^0}^2, m_{{\sb}_m}^2 \Big) 
\\ \non
&& -\vor \sum_{k,m=1}^2 G_{mjk}^{\st\sb} \Big[ h_{b}^2 d^{{\sb}}_{1k} 
\Rst_{i1}\Rst_{m1} + h_{t}^2 d^{{\st}}_{1k} \Rst_{i2}\Rst_{m2} + g^2 
f^{\st}_{im} \Big( d^{\sb}_{1k} - d^{\st}_{1k} \Big) \Big] \times 
\\ \non
&& \hphantom{-\vor \sum_{k,m=1}^2 G_{mjk}^{\st\sb} \Big[} B_0\Big( 
m_{{\sb}_j}^2, m_{H_k^+}^2, m_{{\st}_m}^2 \Big) 
\\ \non
&& -\vor \sum_{k,m=1}^2 G_{imk}^{\st\sb} \Big[ h_{t}^2 d^{{\st}}_{1k} 
\Rsb_{m1}\Rsb_{j1} + h_{b}^2 d^{{\sb}}_{1k} \Rsb_{m2}\Rsb_{j2} + g^2 
f^{\sb}_{mj} \Big( d^{\sb}_{1k} - d^{\st}_{1k} \Big) \Big] \times 
\\
&& \hphantom{-\vor \sum_{k,m=1}^2 G_{mjk}^{\st\sb} \Big[} B_0\Big( 
m_{{\st}_i}^2, m_{H_k^+}^2, m_{{\sb}_m}^2 \Big) 
\end{eqnarray}

\begin{figure}[th]
\begin{picture}(160,215)(0,0)
     \put(0,-2){\mbox{\resizebox{16cm}{!}
     {\includegraphics{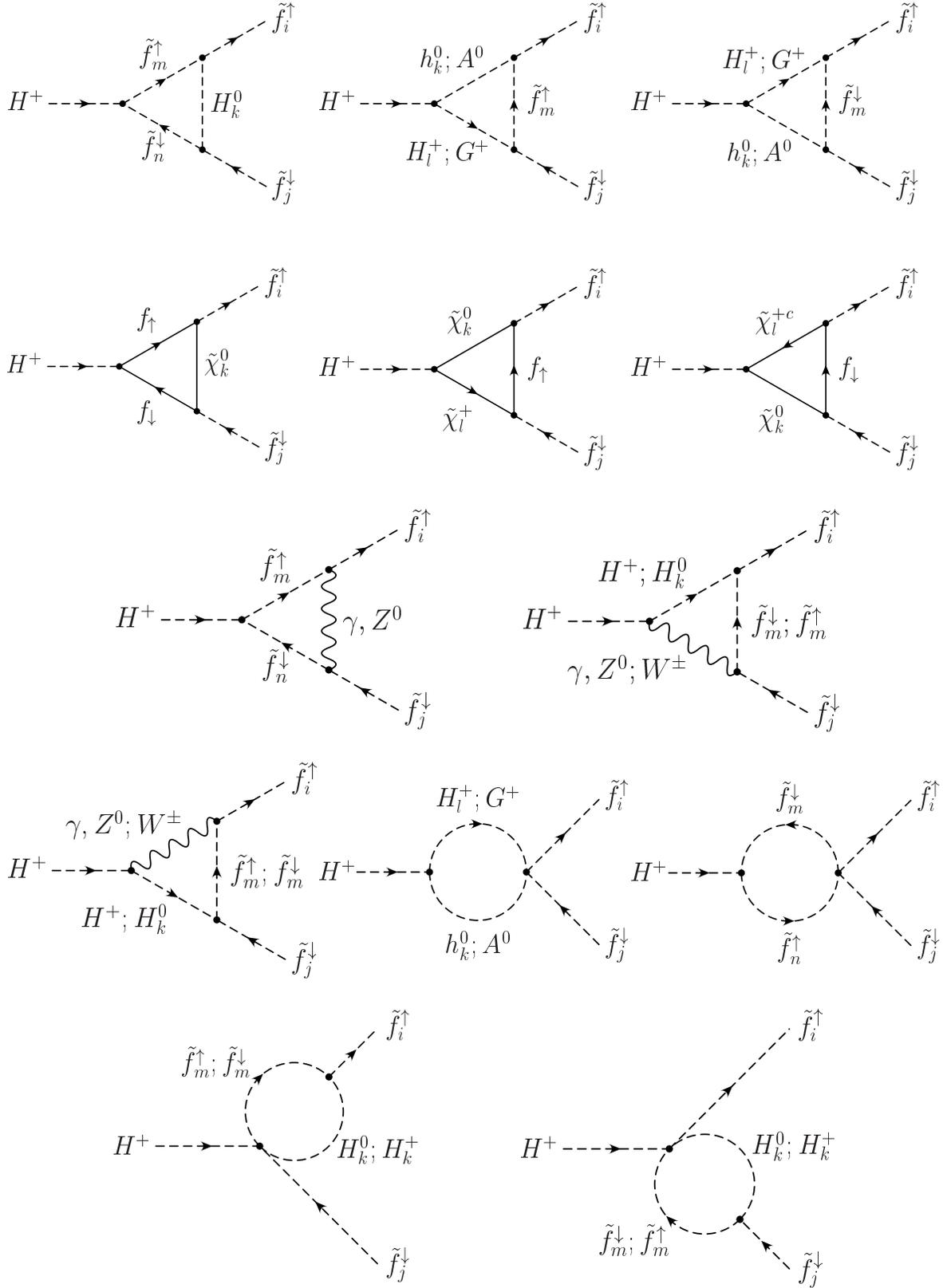}}}}
\end{picture}
\caption{Vertex diagrams relevant to the calculation of the virtual 
electroweak corrections to the decay width $H^+ \rightarrow 
\sf^\uparrow_i \bar{\sf^\downarrow_j}$. In the fourth row, 
$\sf^\uparrow_n$ and $\sf^\downarrow_m$ denote up- and down-type 
sfermions of all three generations, respectively. 
\label{Hpvertex-graphs}} 
\end{figure}

\end{document}